%% file: main.tex
\documentclass[journal=jacsat,manuscript=article,layout=twocolumn]{achemso}

\usepackage[version=3]{mhchem}

\usepackage[fontsize=10pt]{scrextend}
\usepackage{color}
\usepackage{siunitx}
\usepackage{braket}
\usepackage{textgreek}
\usepackage{amssymb}
\usepackage{bm}
\makeatletter
\makeatother

\def\iu{\ensuremath{\mathrm{i}}}
\def\du{\ensuremath{\mathrm{d}}}
\def\eu{\ensuremath{\mathrm{e}}}
\DeclareSIUnit\angstrom{\text{Å}}

\author{Xincheng Miao}
\affiliation[Julius-Maximilians-Universität Würzburg]
{Institut für Physikalische und Theoretische Chemie, Julius-Maximilians-Universität Würzburg, Emil-Fischer-Straße 42, 97074 Würzburg, Germany.}

\author{Toshinori Suzuki}
\email{suzuki@kuchem.kyoto-u.ac.jp}
\affiliation[Kyoto University]
{Department of Chemistry, Graduate School of Science, Kyoto University, Kyoto 606-8502, Japan.}                                                                  

\author{Roland Mitri\'{c}}
\email{roland.mitric@uni-wuerzburg.de}
\affiliation[Julius-Maximilians-Universität Würzburg]
{Institut für Physikalische und Theoretische Chemie, Julius-Maximilians-Universität Würzburg, Emil-Fischer-Straße 42, 97074 Würzburg, Germany.}

\title[]{%
Resolving Photochemical Reaction Pathways Using Field-Induced Surface Hopping Simulation
of fs-EUV-TRPES: The Case of Furan%
}


\let\oldmaketitle\maketitle
\let\maketitle\relax
\SectionNumbersOn

\begin{document}

\begin{tocentry}
    \centering
    \includegraphics[width=3.25in,height=1.75in]{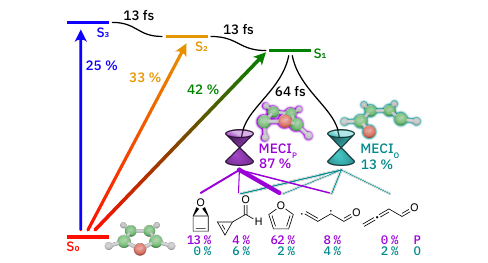}
\end{tocentry}

\twocolumn[
    \begin{@twocolumnfalse}
        \oldmaketitle

        \begin{abstract}
            Recent advances in femtosecond extreme ultraviolet photoelectron spectroscopy
            have enabled unprecedented insight into photochemical processes in the gas
            and condensed phases and at interfaces. However, assigning spectral features
            to specific reaction pathways relies heavily on theoretical signal predictions,
            including a reliable connection between nonadiabatic dynamics and measured observables.
            Here, we present a general approach for resolving photochemical reaction pathways
            by combining field-induced surface hopping (FISH) nonadiabatic dynamics with
            explicit pulse excitation, RMS-CASPT2, and simulations of time-resolved
            photoelectron spectra. The pump laser field prepares the excited-state ensemble
            under experimental conditions, while the dynamically correlated multireference treatment
            describes excited states, conical intersections, and bond-breaking coordinates.
            Mapping trajectories onto photoelectron signals, supported by a systematic analysis
            of photoionization approximations, enables comparison with experiment.
            We illustrate this approach on furan following excitation by a \SI{200}{\nm},
            \SI{10}{\fs} pulse. Among trajectories returning to the ground state, approximately
            \SI{87}{\percent} hop near the ring-puckering conical intersection and \SI{13}{\percent}
            near the ring-opening intersection. Subsequent ground-state vibrational energy redistribution
            nevertheless populates ring-opening structures. Five distinct ground-state products
            are identified. Simulated spectra reproduce time-dependent electron-binding-energy shifts,
            characteristic kinks, and bands with predicted photoproduct contributions.
            Compared with Dyson-norm-based intensities, plane- and Coulomb-wave continuum
            functions improve the early-time shape of the lowest binding energy signal
            and show that photoelectron anisotropy is substantially more sensitive
            to the continuum approximation than the angle-integrated spectrum.
            More generally, these findings demonstrate how explicit-pulse nonadiabatic dynamics
            and dynamically correlated multireference electronic structure, connected to experiment
            through simulated fs-EUV-TRPES, can resolve competing photochemical reaction pathways.
        \end{abstract}

    \end{@twocolumnfalse}
]

\section{Introduction}
Femtosecond extreme-ultraviolet time-resolved photoelectron spectroscopy
(fs-EUV-TRPES) has emerged as a powerful probe of ultrafast molecular
photochemistry, providing simultaneous sensitivity to electronic-state
character, nuclear motion, and product
formation\cite{karashima_ultrafast_2021,karashima_vibrational_2024}. The high
probe-photon energies available in EUV measurements extend the observable
ionization window beyond that accessible with conventional UV probes, enabling
access to multiple ionization channels associated with different neutral and
cationic electronic states along nonadiabatic reaction
pathways\cite{von_conta_conical-intersection_2018}. In gas-phase systems, TRPES
supported by nonadiabatic dynamics simulations has been used to follow
electronic relaxation, conical-intersection dynamics, intersystem crossing,
dissociation, and photochemical product
formation.\cite{mitric_time-resolved_2011,tomasello_exploring_2014,issler_excited_2018,roder_exploring_2019,issler_time-resolved_2023,severino_mapping_2025,karashima_ultrafast_2023}
Its extension to liquid jets has enabled analogous investigations in aqueous
environments\cite{hummert_femtosecond_2018,titov_electronic_2021,kang_dynamics_2024,miura_formation_2023}
and at aqueous interfaces\cite{yamamoto_charge_2026}. Nevertheless, mapping the
measured photoelectron signals onto specific molecular structures or reaction
pathways remains challenging and depends on close integration of experiment and
theory\cite{fielding_using_2018,pitesa_combined_2021_utf,chakraborty_timeresolved_2024}.

Here, we present a general framework for identifying photochemical reaction
pathways from TRPES measurements that combines realistic preparation of the
excited-state ensemble, nonadiabatic dynamics using dynamically correlated
multireference electronic structure methods to describe conical intersections
and bond-breaking coordinates, and simulation of the TRPES observable. Direct
comparison of the simulated and measured signals can then test reaction
pathways derived from the underlying trajectories.

In order to demonstrate this framework, we apply it to furan, which provides a
well-studied yet demanding test case because it combines a rich excited-state
manifold with diverse photochemical
outcomes\cite{fuji_ultrafast_2010,lindner_multistate_2018}. Early gas-phase
absorption measurements delineated the intense $\text{\textpi} \to
    \text{\textpi}^*$ band system in the \SIrange[range-phrase=--]{210}{190}{\nm}
range and revealed weaker features assigned to low-lying Rydberg
states\cite{pickett_vibrational_1940,price_absorption_1941,pickett_vacuum_1951},
while photoelectron and matrix-isolation studies mapped the ionization
landscape and established a chemically diverse S\textsubscript{0} product
inventory including Dewar furan and various open-chain
aldehydes\cite{eland_photoelectron_1969,derrick_photoelectron-spectroscopical_1971,cooper_resonantly_1980,rendall_dewar_1986}.

Although these studies established relevant excitation energies, ionization
potentials and photoproducts, the mechanism by which the deep-UV-excited
population traverses the S\textsubscript{1}/S\textsubscript{0} seam and
subsequently forms hot S\textsubscript{0} products remains unresolved.
State-averaged CASSCF scans identified two minimum-energy conical intersections
(MECIs) on the S\textsubscript{1}/S\textsubscript{0} crossing seam: (i) an
out-of-plane ring-puckering MECI (MECI\textsubscript{P}) and (ii) a planar
ring-opening MECI (MECI\textsubscript{O})\cite{wilsey_role_1996}. Subsequent
scans using DFT/MRCI, EOM-CCSD, and CASPT2 showed that access to
MECI\textsubscript{P} is essentially barrierless and that MECI\textsubscript{O}
is separated by at most a very shallow ($\lesssim \SI{0.1}{\eV}$) barrier,
implying two energetically competitive deactivation
channels\cite{gavrilov_deactivation_2008,gromov_theoretical_2010,stenrup_computational_2010_utf}.

Early femtosecond time-resolved photoelectron spectroscopy established
sub-\SI{100}{\fs} return to S\textsubscript{0} with near-UV
probing\cite{fuji_ultrafast_2010}. Later TRPES with vacuum/extreme-UV probes
followed the spectral evolution from excitation through CI-mediated decay to
hot ground-state products and revealed signatures consistent with relaxation
through
MECI\textsubscript{P}\cite{adachi_probing_2018,uenishi_signatures_2024}. In
parallel, attosecond soft-X-ray absorption and Coulomb-explosion imaging
provided evidence for ultrafast C--O bond
cleavage\cite{severino_attosecond_2024,wang_time-resolved_2023}. Taken
together, these experiments provide evidence for both ring-puckering dynamics
and C--O bond cleavage but do not yield a unified mechanistic picture of the
ultrafast relaxation.

On the theoretical side, nonadiabatic dynamics have likewise yielded
contradictory pictures. Several state-averaged CASSCF surface-hopping studies
predict substantial direct ring opening through
MECI\textsubscript{O}\cite{hua_monitoring_2016,oesterling_substituent_2017_utf}.
In contrast, TDDFT, TDDFTB and ensemble-DFT simulations generally favor
relaxation through
MECI\textsubscript{P}\cite{humeniuk_time-resolved_2013,adachi_probing_2018,filatov_signatures_2021,miron_trajectory_2024},
although other DFT-based studies again predict predominantly ring-opening
dynamics\cite{wang_time-resolved_2023,severino_attosecond_2024}. Consequently,
the relative importance of the two deactivation pathways and their relationship
to the final S\textsubscript{0} product distribution remain uncertain.

To address this unresolved mechanistic question within the framework outlined
above, we investigate the deep-UV photodynamics of furan using field-induced
surface-hopping (FISH) dynamics\cite{mitric_laser-field-induced_2009} combined
with rotated multi-state CASPT2\cite{battaglia_role_2021} electronic structure
calculations. By including the pump field explicitly, a multistate excited
ensemble is prepared under the realistic experimental excitation conditions,
rather than initializing all trajectories in a single excited state, as is
common in field-free fewest-switches surface hopping (FSSH)
simulations\cite{tully_molecular_1990}. The dynamically correlated
multireference treatment provides a balanced description of conical
intersections and bond-breaking dynamics. For a direct connection to
experiment, we compute time-resolved photoelectron spectra along the FISH
trajectories and compare the squared-Dyson-norm approach to photoionization
intensities with explicit matrix elements evaluated with plane- and
Coulomb-wave continuum functions.

We find that ultrafast internal conversion proceeds predominantly through the
puckered S\textsubscript{1}/S\textsubscript{0} intersection, whereas
ring-opened products arise largely from subsequent S\textsubscript{0} dynamics
driven by vibrational energy redistribution into the C--O stretch coordinate.
The simulated TRPES reproduces the rich structure of the experimental
spectrum\cite{uenishi_signatures_2024}, including the main resolved
time-dependent features. This agreement supports tentative assignments of
individual spectral components to specific structural changes connecting the
Franck--Condon region to conical intersections and hot ground-state
photoproducts.

Together, these results provide a unified picture of furan's photodynamics that
reconciles earlier theoretical predictions and rationalizes seemingly
contradictory experimental signatures. In particular, we demonstrate how
evidence for C--O bond cleavage can coexist with predominantly
puckering-mediated internal conversion, consistent with earlier suggestions
that ring opening need not involve direct passage through
MECI\textsubscript{O}\cite{filatov_signatures_2021}. More broadly, this
workflow is transferable to molecules where multi-state excitation, conical
intersections, and competing product channels complicate mechanistic inference
from ultrafast spectra alone.

\section{Methodology}
The computational protocol used in this work was designed to mirror the
sequence of processes probed experimentally, from photoexcitation to
nonadiabatic relaxation and finally photoionization. To this end, we combined
dynamically correlated multireference electronic-structure calculations with
field-induced surface-hopping (FISH) dynamics and simulation of time-resolved
photoelectron spectra. The electronic structure treatment provides a balanced
description of the low-lying valence and Rydberg states involved in the
\SI{200}{\nm} excitation window. An ensemble of molecules initially in the
electronic ground state was subjected to an explicit pump pulse and propagated
during the subsequent relaxation dynamics. Photoelectron observables were
evaluated along the trajectories using ionization energies and
Dyson-orbital-based approximations. The methodological components are outlined
below.

\subsection{Electronic structure and nonadiabatic dynamics}

\begin{figure}
    \centering
    \includegraphics[width=\linewidth]{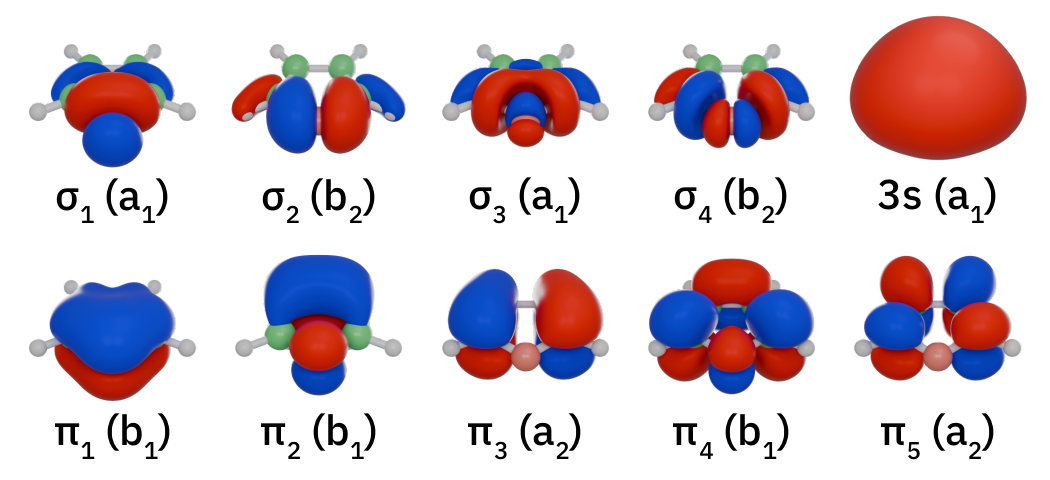}
    \caption{%
        Natural orbitals of furan at the SA5-RMS-CASPT2(10,10)/aug-cc-pVDZ level.
        An isovalue of 0.04 is used for the valence orbitals
        and 0.01 for the Rydberg orbital.
        The 3s Rydberg orbital is displayed at a reduced scale for clarity.%
    }
    \label{fig:furan_nos}
\end{figure}

We describe the electronic structure of furan using the state-averaged complete
active space self-consistent field (SA-CASSCF)
method\cite{werner_quadratically_1981} in combination with the rotated
multi-state second-order perturbation theory
(RMS-CASPT2)\cite{battaglia_role_2021} to include dynamical correlation
effects, as implemented in the \textsc{OpenMolcas} program
package\cite{li_manni_openmolcas_2023}. The RMS variant is chosen for its
better root-ordering when state mixing occurs. An imaginary shift of
\SI{0.2}{{\textit{E}_h}} was used to mitigate intruder
states\cite{forsberg_multiconfiguration_1997_utf}. No ionization
potential--electron affinity shift was applied. To treat the diffuse Rydberg
states in furan, the aug-cc-pVDZ basis set\cite{kendall_electron_1992} was
employed for CASPT2 calculations. We chose an active space of 10 electrons in
10 orbitals, comprising the five $\pi$ and $\pi^*$ orbitals, the 3s Rydberg
orbital, and the two $\sigma_{\text{CO}}$ and the two $\sigma_{\text{CO}}^*$
orbitals needed to describe the CO bond elongation and rupture on the lowest
excited singlet surface\cite{fuji_ultrafast_2010,schalk_ring-opening_2018}.
State averaging was performed over five singlet states, comprising the $1\
    ^1A_1$ ground state, the $1\ ^1A_2$ 3s Rydberg state, the $1\ ^1B_2$ and $2\
    ^1A_1$ valence-excited states, and an additional $B_1$ state. Notice that at
the CASSCF level, the $2\,^1A_1$ state is not among the four lowest singlet
states. This SA5-RMS-CASPT2(10, 10)/aug-cc-pVDZ treatment was adopted because,
near the ground-state equilibrium geometry, the $1\ ^1B_2$ is not among the
lowest four roots at the SA-CASSCF level, and a $B_1$ state enters instead.
Upon inclusion of dynamic correlation, the physically relevant low-lying state
manifold is restored, and only the first four states are considered in the
following. The corresponding natural orbitals are shown in
Fig.~\ref{fig:furan_nos}.

The nonadiabatic dynamics simulations were performed using the
FISH\cite{mitric_laser-field-induced_2009,mitric_field-induced_2011} procedure
combined with the RMS-CASPT2 electronic structure method, which allows the
inclusion of explicit laser fields. The nuclear motion is treated classically
using Newtonian equations, while the electronic dynamics is simulated by
solving the time-dependent Schrödinger equation, given by
\begin{align}
    \label{eq:fish_equation}
    \iu \hbar \dot{c}_i(t) & = E_{i}(\vec{R}(t))c_i(t) - \iu \hbar \sum_j D_{ij} (\vec{R}(t))c_j(t) \nonumber                    \\
                           & \phantom{=\ } - \sum_j \left( \vec{\mu}_{ij} (\vec{R}(t)) \cdot \vec{\varepsilon}(t) \right) c_j(t)
\end{align}
where $c_i(t)$ represents the expansion coefficient of the electronic wavefunction
in terms of the $i$-th electronic state and $E_i$ is the adiabatic energy of the electronic state,
whereas $D_{ij}$ stands for the scalar nonadiabatic coupling between the electronic states
$i$ and $j$, given by
\begin{equation}
    D_{ij} (\vec{R}(t)) = \underbrace{\braket{\psi_i (\vec{R}(t))|\nabla_R|\psi_j (\vec{R}(t))}}_{\vec{d}_{ij}} \cdot \frac{\du  \vec{R}(t)}{\du t}\,.
\end{equation}
The scalar product $\vec{\mu}_{ij}\cdot \vec{\varepsilon} (t)$ describes the interaction
of the molecular transition dipole moment $\vec{\mu}_{ij}$ connecting the electronic states
$i$ and $j$ with the external laser field $\vec{\varepsilon}(t)$.
In the simulation, the laser pulse is modeled as a
Gaussian envelope with a carrier frequency of \SI{6.199}{\eV}
(\SI{200}{\nm}), an intensity FWHM of \SI{10}{\fs}, and an electric field
strength of \SI{0.01}{{a.u.}} These parameters approximate the
\SI{200}{\nm} pump pulse used by Uenishi \emph{et al.}\cite{uenishi_signatures_2024}

The transition probability from state $i$ to state $j$ is expressed as the
product of two factors: the probability of depopulating state $i$
($P_{i,\text{depopulation}}$) and the probability of populating state $j$
($P_{j,\text{population}}$), calculated as
follows\cite{lisinetskaya_simulation_2011}:
\begin{subequations}
    \label{eq:sh_probabilities}
    \begin{align}
        P_{i,\text{depopulation}} & = \Theta(-\dot{\rho}_{ii}) \frac{-\dot{\rho}_{ii}}{\rho_{ii}} \Delta t                            \\
        P_{j,\text{population}}   & = \frac{\Theta(\dot{\rho}_{jj})\dot{\rho}_{jj}}{\sum_k \Theta(\dot{\rho}_{kk})\dot{\rho}_{kk}}\,,
    \end{align}
\end{subequations}
where $\Theta$ is the Heaviside step function.

For dynamics simulations, the adiabatic energies were obtained at the
SA5-RMS-CASPT2(10, 10)/aug-cc-pVDZ level of theory with nonadiabatic coupling
vectors\cite{nishimoto_analytic_2022} calculated among the first four states.
The initial structures were generated by sampling a harmonic canonical Wigner
distribution at 50 K. Since no floppy low-frequency modes are present (lowest
frequency \SI{575}{\per\cm}), this approach is expected to produce a reasonable
approximation of the quantum mechanical distribution.

Because the S\textsubscript{1} and S\textsubscript{3} states becomes bright at
distorted geometries, we include a laser pulse explicitly in the dynamics
simulations to generate a physically reasonable excited-state population
distribution, in contrast to the frequently used approach of promoting all the
trajectories to the brightest S\textsubscript{2} state. A total of 250
trajectories were subjected to the laser pulse, and 55 of them were found to be
in an excited state after the pulse. These trajectories were propagated for an
additional \SI{200}{\fs} with the field set to zero. The nuclear time step was
set to \SI{0.2}{\fs}. Since the dynamics proceeds sequentially, no decoherence
correction was applied, as the relaxation process happens sequentially without
oscillations.

\subsection{Simulation of photoelectron spectra}

\subsubsection*{Cross sections and Dyson Orbitals}

Throughout this subsection, Hartree atomic units are used. Within the
electric-dipole approximation and first-order perturbation theory, the
differential photoionization cross section for ionization from an initial
$N$-electron state $\varPsi_I^N$ into a final ionic state $\varPsi_F^{N-1}$
with an outgoing electron of momentum $\vec{k}$ can be written as
\begin{equation}
    \label{eq:general_photoelectron_differential_cross_section}
    \frac{\du \sigma_{IF}}{\du \Omega_{\hat k}} = \frac{4\pi^2}{c} k E g_F \left| D_{IF}(\vec{k}) \right|^2,
\end{equation}
where $\Omega_{\hat k}$ denotes the solid angle of the photoelectron emission
direction $\hat{k}=\vec{k}/k$, $c$ is the speed of light in vacuum,
$k = \| \vec{k} \|$ is the magnitude of the photoelectron wave vector, and
$E$ is the photon energy.
Furthermore, $g_F$ is the degeneracy of the final ionic channel,
and $D_{IF}(\vec{k})$ is the photoelectron dipole matrix element,
given by
\begin{equation}
    \label{eq:general_photoelectron_dipole_moment}
    D_{IF}(\vec{k}) = \Braket{ \hat{\mathcal{A}}
        \left[ \varPsi_F^{N-1} \psi_{\vec{k}}^{(-)} \right] | \vec{\varepsilon} \cdot \hat{\vec{\mu}} | \varPsi_I^N }\,.
\end{equation}
The operator $\hat{\mathcal{A}}$ denotes normalized antisymmetrization of the final
$N$-electron scattering state.
The continuum states are momentum-normalized,
$\braket{\psi_{\vec{k}}|\psi_{\vec{k}'}}=\delta^{(3)}(\vec{k}-\vec{k}')$,
where $\delta^{(3)}$ denotes the three-dimensional Dirac delta distribution.
The quantities $\hat{\vec{\mu}}$, $\vec{\varepsilon}$,
and $\psi_{\vec{k}}^{(-)}$ denote the dipole operator,
the light-polarization vector, and the continuum wave function of the
emitted electron, respectively.

Assuming strong orthogonality between the continuum orbital and the bound
orbitals, the many-electron photoionization matrix element can be reduced to a
one-electron transition amplitude involving the Dyson orbital
$\phi^{\mathrm{D}}_{IF}$:\cite{oana_dyson_2007,oana_cross_2009}
\begin{equation}
    \label{eq:photoelectron_dipole_moment_dyson}
    D_{IF}(\vec{k}) = \braket{ \psi_{\vec{k}}^{(-)} | \vec{\varepsilon} \cdot \hat{\vec{\mu}} | \phi^{\mathrm{D}}_{IF} }
\end{equation}
with
\begin{align}
    \label{eq:dyson_orbital_definition}
    \phi^{\mathrm{D}}_{IF}(\vec{r}_1)
    =
    \sqrt{N}
     & \int \varPsi_F^{N-1}(x_2,\cdots,x_N)^* \nonumber \\
     & \times \varPsi_I^N(x_1,x_2,\cdots,x_N)
    \, \du\sigma_1\du x_2 \cdots \du x_N\,,
\end{align}
where $x_i=(\vec{r}_i,\sigma_i)$ denotes the combined spatial and spin coordinates
of the $i$-th electron, and integration over $\sigma_1$ denotes summation over
the spin coordinate of the first electron.

The Dyson orbital therefore plays the role of the effective one-electron
orbital associated with ionization from $\varPsi_I^N$ to $\varPsi_F^{N-1}$,
whereas $\psi_{\vec{k}}^{(-)}$ describes the outgoing photoelectron in the
field of the ion. In the simulation, the Dyson orbitals are computed from
neutral and cationic wavefunctions obtained at the SA5-RMS-CASPT2(10, 10) and
SA8-RMS-CASPT2(9, 10) levels, respectively.

With the Dyson orbitals fixed by the chosen electronic-structure method, the
approximation considered here concerns the continuum wave function
$\psi_{\vec{k}}^{(-)}$. In a single-center representation, the scattering state
is expanded in partial waves
\begin{equation}
    \label{eq:partial_wave_expansion}
    \psi_{\vec{k}}^{(-)}(\vec{r})
    = \frac{4\pi}{(2\pi)^{3/2}}
    \sum_{l=0}^{\infty} \sum_{m=-l}^{l}
    \iu^l\, R_{kl}(r)\, Y_{lm}(\hat{r})\,Y_{lm}^{*}(\hat{k})\,,
\end{equation}
where the angular dependence, specified by the unit vector
$\hat r=\vec r/r$, is carried by spherical harmonics $Y_{lm}$,
while the choice of continuum approximation enters through the radial functions
$R_{kl}(r)$\cite{sakurai_modern_1994}.

For a neutral molecular core, for example in anion photodetachment, the
continuum electron experiences asymptotically no long-range Coulomb potential.
The continuum therefore approaches a free-electron form asymptotically, which
is obtained by choosing the regular radial solutions $R_{kl}(r)=j_{l}(kr)$,
where $j_l$ is the spherical Bessel function of the first kind. These solutions
recover the plane-wave form
\begin{equation}
    \label{eq:plane_wave}
    \psi_{\vec{k}}^{(-)}(\vec{r})
    =
    \frac{1}{(2\pi)^{3/2}}
    \eu^{\iu\vec{k}\cdot\vec{r}}\,.
\end{equation}
This plane-wave form neglects the short-range scattering
potential of the residual core and is therefore most appropriate as a
high-kinetic-energy approximation.

For a charged molecular core, for example in photoionization, the outgoing
electron experiences a long-range Coulomb potential. The corresponding regular
radial solutions are the regular Coulomb functions $F_l(\eta,\rho)$, entering
the partial-wave expansion for the $(-)$ boundary condition through
$R_{kl}(r)=\exp(-\iu\sigma_l)F_l(\eta,kr)/(kr)$, where
$\sigma_l=\arg\Gamma(l+1+\iu\eta)$, with $\Gamma$ denoting the gamma function,
is the Coulomb phase shift and $\eta=-Z/k$ is the Sommerfeld parameter for a
residual charge $Z$ in atomic units. This Coulomb-wave form therefore gives the
correct long-range asymptotics and provides a more appropriate
high-kinetic-energy approximation than the plane-wave form. In the
high-kinetic-energy limit at fixed radial coordinate, $k\to\infty$ and hence
$\eta\to 0$. The regular Coulomb radial function then reduces to the spherical
Bessel function\cite{olver_nist_2026}, so that the Coulomb- and plane-wave
approximations become equivalent. This statement concerns the high-energy limit
and should not be confused with the large-distance asymptotic limit
$r\to\infty$, where the Coulomb phase remains relevant.

A commonly used further simplification is to replace the full matrix element by
the squared Dyson norm $\|\phi^{\mathrm D}_{IF}\|^2$, which serves as a
qualitative estimate of the relative ionization
strength\cite{diaztinoco_dyson_2019}. This approximation retains information
about the overlap between the neutral and ionic wave functions, but it neglects
the explicit continuum electron and therefore omits the energy, angular, and
polarization dependence of the true photoionization matrix element.

It should be noted that while the plane-wave and Coulomb-wave approximations
deliver the proper high-kinetic-energy limit for the continuum description, the
squared Dyson norm does not, in general, constitute a controlled asymptotic
approximation to relative photoionization intensities. In order to illustrate
this point, consider two ionization channels with normalized Gaussian Dyson
orbitals
\begin{equation}
    \label{eq:example_two_channel_dyson}
    \phi_1(\vec{r}) = \left(\frac{2\alpha_1}{\pi}\right)^{3/4} \eu^{-\alpha_1 r^2},\quad
    \phi_2(\vec{r}) = \left(\frac{2\alpha_2}{\pi}\right)^{3/4} \eu^{-\alpha_2 r^2},
\end{equation}
with exponents $\alpha_1\neq \alpha_2$.
By construction, one has $\|\phi_1\|^2=\|\phi_2\|^2=1$,
so the Dyson-norm estimate predicts a relative intensity ratio of
$\|\phi_1\|^2 / \|\phi_2\|^2 = 1$.

Using the plane-wave approximation, one can apply
Eq.~\eqref{eq:photoelectron_dipole_moment_dyson} to obtain
\begin{equation}
    \label{eq:example_two_channel_dipole_moment}
    D_s(\vec{k}) = -\frac{\iu \vec{\varepsilon} \cdot \vec{k}}{(2\pi\alpha_s)^{3/4}}
    \frac{1}{2\alpha_s} \eu^{-\frac{k^2}{4\alpha_s}}
\end{equation}
for $s=1,2$. By applying
Eq.~\eqref{eq:general_photoelectron_differential_cross_section}
using fixed polarization $\vec{\varepsilon}=\hat{e}_z$, where $\hat{e}_z$ is the unit vector
along the $z$ axis and $k_z = \vec{k}\cdot\hat{e}_z$, one obtains
\begin{equation*}
    \frac{\du \sigma_s^{[z]}}{\du \Omega_{\hat k}}
    = \frac{4\pi^2}{c}kE \frac{k_z^2}{4\alpha_s^2 (2\pi\alpha_s)^{3/2}} \eu^{-\frac{k^2}{2\alpha_s}}
\end{equation*}
after inserting Eq.~\eqref{eq:example_two_channel_dipole_moment}.
Integrating over all solid angles, one obtain the total cross section
\begin{equation*}
    \sigma_s^{[z]} = \frac{16\pi^3}{3c}k^3E \frac{1}{4\alpha_s^2 (2\pi\alpha_s)^{3/2}} \eu^{-\frac{k^2}{2\alpha_s}}\,.
\end{equation*}
Since the Dyson orbitals are spherically symmetric, the isotropically
averaged total cross section $\sigma_s$ will be the same as $\sigma_s^{[z]}$,
leading to the intensity ratio
\begin{equation*}
    \frac{\sigma_1}{\sigma_2} = \left(\frac{\alpha_2}{\alpha_1}\right)^{7/2}
    \exp{\left[ -\frac{k^2}{2} \left( \frac{1}{\alpha_1} - \frac{1}{\alpha_2} \right) \right]}\,.
\end{equation*}
It is clear that this ratio does not equal to 1 in general.
In the high-kinetic-energy limit, $k\to\infty$,
it approaches either 0 or $\infty$ for $\alpha_1 \neq \alpha_2$,
rather than the unit ratio predicted from the squared Dyson norms.
This simple example illustrates that squared Dyson norms provide
only a qualitative estimate of relative ionization intensities.
In addition to the radial form of the continuum, its single-center representation
requires a choice of origin.

\subsubsection*{Choice of the origin}

The choice of the origin of this single-center representation is not unique.
Two physically motivated choices are the Dyson centroid and the charge center
of the residual ionic potential. The Dyson centroid is defined as
\begin{equation}
    \label{eq:dyson_centroid}
    \vec{R}^{\mathrm D}_{IF}
    = \frac{
        \int \vec{r}\, |\phi^{\mathrm D}_{IF}(\vec r)|^2 \,\du^3\vec{r}
    }{
        \int |\phi^{\mathrm D}_{IF}(\vec r)|^2 \, \du^3\vec{r}
    }\,.
\end{equation}
It is a property of the ionization transition amplitude and therefore depends on both the initial neutral state and the final ionic state. By contrast, the charge center is defined by
\begin{equation}
    \label{eq:charge_center_definition}
    \vec R^{\mathrm Q}_{F}
    = \frac{
        \int \vec{r}\, \rho_F^{\mathrm{ion}}(\vec{r})\, \du^3\vec{r}
    }{
        \int \rho_F^{\mathrm{ion}}(\vec{r})\, \du^3\vec{r}
    },.
\end{equation}
Here, the ionic charge density is
\begin{equation}
    \label{eq:ionic_charge_density}
    \rho_F^{\mathrm{ion}}(\vec{r})
    =
    \rho^{\mathrm{nuc}}(\vec{r})
    -
    \rho_F^{N-1}(\vec{r}),.
\end{equation}
The nuclear charge density is given by
\begin{equation}
    \label{eq:nuclear_charge_density}
    \rho^{\mathrm{nuc}}(\vec{r})
    =
    \sum_A Z_A\delta^{(3)}(\vec{r}-\vec{R}_A)
\end{equation}
with nuclear charges $Z_A$ at positions $\vec{R}_A$.
The cationic electron density $\rho_F^{N-1}(\vec r)$ is normalized to $N-1$ electrons,
namely
\begin{equation}
    \int \rho_F^{N-1}(\vec r)\, \du^3\vec r = N-1\,.
\end{equation}
The denominator of Eq.~\eqref{eq:charge_center_definition} is therefore the total ionic charge $Q_F^{N-1}$, and its numerator is the dipole moment of the cation $\vec{\mu}_F^{N-1}$, leading to
\begin{equation}
    \label{eq:charge_center_dipole_relation}
    \vec R^{\mathrm Q}_{F}
    =
    \frac{\vec{\mu}_F^{N-1}}{Q_F^{N-1}},.
\end{equation}

The two centers are generally different, even in the simplest
single-determinant, Koopmans-like ionization scheme. Let $\rho_I^N(\vec r)$
denote the one-electron density of the neutral initial state, normalized to
$N$. In a single-determinant description,
\begin{equation}
    \label{eq:neutral_density_koopmans}
    \rho_I^N(\vec r)
    =
    \sum_{i=1}^{n_{\mathrm{occ}}}
    |\varphi_i(\vec r)|^2
\end{equation}
with $n_{\mathrm{occ}}$ occupied spin-orbitals $\varphi_i$.
If ionization removes the electron from the occupied spin-orbital $\varphi_j$,
the cationic density is, in the Koopmans approximation,
\begin{equation}
    \label{eq:cation_density_koopmans}
    \rho_F^{N-1}(\vec r)
    =
    \sum_{i\neq j}^{n_{\mathrm{occ}}}
    |\varphi_i(\vec r)|^2
    =
    \rho_I^N(\vec r)-|\varphi_j(\vec r)|^2\,.
\end{equation}
Consequently,
\begin{equation}
    \label{eq:ionic_charge_density_koopmans}
    \rho_F^{\mathrm{ion}}(\vec r)
    =
    \rho^{\mathrm{nuc}}(\vec r)
    -
    \rho_I^N(\vec r)
    +
    |\varphi_j(\vec r)|^2\,.
\end{equation}
Inserting this into Eq.~\eqref{eq:charge_center_definition}, one obtains
\begin{align*}
    \vec{R}^{\mathrm{Q}}_F & = \frac{
        \int \vec{r}\, \left[
            \rho^{\mathrm{nuc}}(r) - \rho_I^N(\vec{r}) + |\varphi_j(\vec{r})|^2
            \right]\, \du^3\vec{r}
    }{Q_F^{N-1}}                                                                                                         \\
                           & = \frac{
        \int \vec{r}\, \left[
            \rho^{\mathrm{nuc}}(r) - \rho_I^N(\vec{r})
            \right]\, \du^3\vec{r} + \int \vec{r}\, |\varphi_j(\vec{r})|^2\, \du^3\vec{r}
    }{Q_F^{N-1}}                                                                                                         \\
                           & = \frac{\vec{\mu}_I^N + \int \vec{r}\, |\varphi_j(\vec{r})|^2\, \du^3\vec{r}}{Q_F^{N-1}}\,,
\end{align*}
where $\vec{\mu}_I^N$ is the dipole moment of the $N$-electron species.
On the other hand, the Dyson orbital is just $\varphi_j(\vec{r})$ in this approximation,
one has $\vec{R}^{\mathrm{D}}_{IF}=\int \vec{r}\, |\varphi_{j}(\vec{r})|^2\, \du^3 \vec{r}$.
The denominator in Eq.~\eqref{eq:dyson_orbital_definition} is 1, since the
molecular orbital is normalized.
Therefore, we obtain the relation
\begin{equation}
    \label{eq:dyson_centroid_charge_center_relation}
    \vec{R}^{\mathrm{Q}}_F = \frac{\vec{R}^{\mathrm{D}}_{IF} + \vec{\mu}_I^N}{Q_F^{N-1}}\,.
\end{equation}
Beyond the Koopmans picture this relation is further modified by orbital relaxation
and correlation.

The Dyson centroid has been established as a convenient single-center origin in
Dyson-orbital calculations of photoionization and photodetachment
crosssections.\cite{oana_cross_2009,gozem_photoelectron_2015} Its main
advantage can be understood from the relation between Dyson centering and
Gram--Schmidt orthogonalization of the approximate continuum wave function. Let
an approximate continuum state be orthogonalized to the Dyson orbital according
to
\begin{equation}
    \label{eq:orthogonalized_continuum_dyson}
    \ket{\psi_{\vec{k}}^{\perp}}
    =
    \ket{\psi_{\vec{k}}}
    -
    \ket{\phi^{\mathrm{D}}_{IF}}
    \frac{
        \braket{\phi^{\mathrm{D}}_{IF}|\psi_{\vec{k}}}
    }{
        \braket{\phi^{\mathrm{D}}_{IF}|\phi^{\mathrm{D}}_{IF}}
    }\,.
\end{equation}
The corresponding length-gauge dipole matrix element is
\begin{align}
    \label{eq:orthogonalized_dipole_matrix_element}
    D_{IF}^{\perp}(\vec{k})
     & = \braket{\psi_{\vec{k}}^{\perp} | \vec\varepsilon\cdot\vec r | \phi^{\mathrm{D}}_{IF}} \nonumber \\
     & = \braket{\psi_{\vec{k}} | \vec\varepsilon\cdot\vec r | \phi^{\mathrm{D}}_{IF}}
    - \frac{\braket{\psi_{\vec{k}}|\phi^{\mathrm{D}}_{IF}}}{\braket{\phi^{\mathrm{D}}_{IF}|\phi^{\mathrm{D}}_{IF}}}
    \braket{\phi^{\mathrm{D}}_{IF} | \vec\varepsilon\cdot\vec{r} | \phi^{\mathrm{D}}_{IF}}\,.
\end{align}
If the coordinate origin is placed at the Dyson centroid,
\begin{equation}
    \braket{\phi^{\mathrm{D}}_{IF} | \vec{r} | \phi^{\mathrm{D}}_{IF}} = \vec{0}\,,
\end{equation}
and therefore the second term in Eq.~\eqref{eq:orthogonalized_dipole_matrix_element} vanishes.
Thus, in the length gauge, evaluating the transition amplitude with a non-orthogonalized
continuum wave about the Dyson centroid gives the same matrix element as evaluating it
with a continuum wave explicitly orthogonalized to the Dyson orbital.
This is the sense in which the Dyson-centered convention implicitly accounts for the
leading orthogonalization correction.

Orthogonalized plane waves have a long history in molecular and atomic
photoionization
calculations.\cite{ellison_theoretical_1974,rabalais_calculated_1974,deleuze_plane_1994,williams_comparison_1977}
They remove the spurious component of a plane-wave-like continuum in the bound
orbital space and thereby improve the formal consistency of the transition
amplitude, especially in the length gauge where such overlap enters through the
position operator. This correction, however, only addresses the orthogonality
defect of the approximate continuum and does not turn the plane wave into a
true scattering solution. Accordingly, comparisons with Hartree--Fock or
scattering-state continua show that orthogonalized waves can still give
inaccurate absolute intensities and angular
distributions.\cite{williams_comparison_1977,deleuze_plane_1994}

For plane waves, changing the single-center origin does not change the physical
continuum model. If the origin is shifted by a vector ($\vec{a}$), the plane
wave acquires only a global phase factor,
\begin{equation}
    \exp[\iu\vec k\cdot(\vec r+\vec a)]
    = \exp(\iu\vec k\cdot\vec a) \exp(\iu\vec k\cdot\vec r),
\end{equation}
which does not affect cross sections. The choice of origin is therefore mainly a matter of
the matrix-element representation and of the convergence of the single-center expansion.

For Coulomb waves the situation is different, because shifting the origin also
shifts the center of the model Coulomb potential. The Dyson center optimizes
the length-gauge matrix element with respect to orthogonalization to the Dyson
orbital, but it does not necessarily give the best single-center representation
of the long-range ionic potential.

The charge center $\vec{R}_F^{\mathrm Q}$ instead follows from the far-field
potential of the residual ion. With respect to a chosen origin, the
electrostatic potential energy of the outgoing electron has the asymptotic form
\begin{equation}
    \label{eq:ionic_potential_multipole}
    V_F(\vec r)
    =
    -\frac{Q_F^{N-1}}{r}
    -
    \frac{\vec \mu_F^{N-1}\cdot\hat r}{r^2}
    +
    \mathcal O(r^{-3}),
\end{equation}
up to the sign convention used for the molecular dipole moment. If the origin is shifted to
the charge center $\vec{R}_F^{\mathrm Q}=\vec{\mu}_F^{N-1}/Q_F^{N-1}$, the dipole term of the
ionic charge distribution vanishes, and the leading correction to the central Coulomb potential
becomes quadrupolar, $\mathcal{O}(r^{-3})$.
A charge-centered Coulomb wave is therefore a natural one-center approximation to the
long-range ionic field. Unlike for a Dyson-centered wave, however, the cancellation in
Eq.~\eqref{eq:orthogonalized_dipole_matrix_element} does not generally occur because the
Dyson-orbital dipole moment about \(\vec{R}_F^{\mathrm Q}\) is nonzero.
In charge-centered Coulomb-wave calculations, the continuum should therefore be explicitly orthogonalized
to the Dyson orbital.
Related approximate continuum models have been used in
time-resolved photoelectron simulations,\cite{tao_ultrafast_2011} and effective-charge
or multicenter Coulomb-wave variants have been introduced to improve the representation of
the residual ionic field beyond a single unit central charge.\cite{gozem_photoelectron_2015,ru_dipole_2022}

\subsubsection*{Rotational averaging}

The photoionization cross section defined in
Eq.~\eqref{eq:general_photoelectron_differential_cross_section} applies to a
fixed molecular orientation in the laboratory frame. In experiments, however,
the molecular orientations are in general not fixed, and the corresponding
cross section must therefore be averaged over the relevant orientational
distribution. To this end, the Dyson orbital and continuum wave function are
represented in a single-center expansion about the chosen origin, either
$\vec{R}^{\mathrm{D}}_{IF}$ or $\vec{R}^{\mathrm{Q}}_{F}$.

The matrix element in Eq.~\eqref{eq:photoelectron_dipole_moment_dyson} is then
evaluated in a single-center partial-wave representation. For each outgoing
partial wave and dipole component, the angular dependence of the
dipole-weighted Dyson orbital is projected onto spherical harmonics, yielding
amplitudes of the form
\begin{equation}
    \label{eq:partial_wave_dipole_amplitudes}
    D_{l m \mu}
    =
    \iu^l
    \int_0^\infty r^3 R_{kl}^{*}(r)
    \int \phi^{\mathrm{D}}_{IF}(r,\hat{r}) Y_{lm}(\hat{r}) Y_{1\mu}(\hat{r})
    \, \du\Omega_{\hat r}\, \du r\,.
\end{equation}
up to the phase and normalization convention used for the spherical harmonics.
Here, $\mu=-1,0,1$ labels the spherical component of the dipole operator,
and $\Omega_{\hat r}$ denotes the solid angle associated with the
spatial direction $\hat r$.

When explicit orthogonalization of the continuum to the Dyson orbital is used,
these amplitudes are corrected before the rotational average is performed.
Defining the continuum--Dyson overlap
\begin{equation}
    \label{eq:partial_wave_overlap}
    S_{lm} = \iu^l \int_0^\infty r^2 R_{kl}^{*}(r) \int \phi^{\mathrm{D}}_{IF}(r, \hat{r})Y_{lm}(\hat{r})
    \,\du\Omega_{\hat r}\,\du r
\end{equation}
and the spherical components of the Dyson-orbital dipole moment
\begin{equation}
    \label{eq:dyson_spherical_dipole_moment}
    M_\mu = \int_0^\infty r^3 \int |\phi^{\mathrm{D}}_{IF}(r, \hat r)|^2 Y_{1\mu}(\hat{r})
    \,\du\Omega_{\hat r}\,\du r\,,
\end{equation}
Gram--Schmidt orthogonalization leads to
\begin{equation}
    \label{eq:orthogonalized_partial_wave_dipole_amplitudes}
    D^{\perp}_{lm\mu} = D_{lm\mu} - S_{lm} \frac{M_\mu}{\|\phi^{\mathrm{D}}_{IF}\|^2}\,.
\end{equation}
For Dyson-centered calculations, $M_\mu = 0$, and therefore $D^{\perp}_{lm\mu} = D_{lm\mu}$.
For charge-centered calculations, $M_\mu \neq 0$ in general, and the orthogonalized amplitudes
$D^{\perp}_{lm\mu}$ are used in the rotational average.
In the following, $\tilde D_{lm\mu}=D_{lm\mu}$ denotes non-orthogonalized
or Dyson-centered amplitudes, whereas $\tilde D_{lm\mu}=D^{\perp}_{lm\mu}$
denotes explicitly orthogonalized amplitudes.

Consider a molecular ensemble whose orientational distribution is axially
symmetric about the laboratory-frame axis $\hat e$. After averaging over
molecular orientations, the photoelectron angular distribution depends only on
the angle $\theta=\arccos(\hat k\cdot\hat e)$ between the photoelectron
direction and the symmetry axis. It can therefore be expanded as
\begin{equation}
    \label{eq:rotationally_averaged_dipole}
    \left\langle
    \left|D_{IF}(\vec k)\right|^2
    \right\rangle
    =
    \sum_{L=0}^{\infty} b_L P_L(\cos\theta)\,,
\end{equation}
where $\langle\cdot\rangle$ denotes averaging over molecular orientations at
fixed photoelectron momentum magnitude $k$, and $P_L$ is the $L$-th Legendre
polynomial.

Conversely, with $x=\cos\theta$, the coefficients can be obtained by projection
\begin{equation}
    \label{eq:legendre_coefficient_projection}
    b_L
    =
    \frac{2L+1}{2}
    \int_{-1}^{1}
    \left\langle
    \left|D_{IF}(\vec k)\right|^2
    \right\rangle
    P_L(x)\,\du x\,.
\end{equation}

The total cross section is determined by the zeroth Legendre coefficient. With
the help of Eq.~\eqref{eq:general_photoelectron_differential_cross_section},
one obtains
\begin{align}
    \label{eq:total_cross_section_from_b0}
    \sigma_{IF}
     & =
    \frac{4\pi^2}{c} kE g_F
    \int
    \left\langle
    \left|D_{IF}(\vec k)\right|^2
    \right\rangle
    \,\du\Omega_{\hat k}
    \nonumber \\
     & =
    \frac{16\pi^3}{c} kE g_F\, b_0\,.
\end{align}
The photoelectron anisotropy parameters are defined by
\begin{equation}
    \label{eq:beta_definition}
    \beta_{L,IF}
    =
    \frac{b_L}{b_0}.
\end{equation}

To describe the orientational distribution, let $\hat a$ be a reference
direction fixed in the molecular frame, and let $\mathcal R$ denote the
rotation from the molecular to the laboratory frame. The angle
\begin{equation}
    \vartheta
    =
    \arccos\!\left[
        \bigl(\mathcal R\hat a\bigr)\cdot\hat e
        \right]
\end{equation}
specifies the orientation of this molecular direction relative to the
symmetry axis. For an axial orientational distribution $W(\vartheta)$, the
Legendre coefficients can be obtained from the corresponding
channel amplitudes $\tilde D_{lm\mu}$ \emph{via}
\begin{equation}
    \label{eq:axial_legendre_coefficients}
    b_L
    =
    \sum_{l m \mu}
    \sum_{l' m' \mu'}
    C^{L}_{l m \mu,l' m'\mu'}[W](p)
    \tilde D_{lm\mu}
    \tilde D_{l'm'\mu'}^{*}\,,
\end{equation}
where $p$ denotes the spherical polarization component of the probe pulse,
with $p=0$ being linear polarization parallel to $\hat e$
and $p=\pm 1$ being circular polarization about $\hat e$.
The rotational-averaging coefficients are\cite{underwood_time-resolved_2000,hockett_general_2015}
\begin{equation}
    \label{eq:axial_rotational_averaging_coefficients}
    \begin{aligned}
        C^{L}_{l m \mu,l' m'\mu'}[W](p)
         & =
        \sum_{K=0}^{2}
        \frac{(2K+1)(2L+1)}{4\pi}
        \\
         & \hspace{-1ex}{}\times
        (-1)^{\mu+p+m'}
        \sqrt{(2l+1)(2l'+1)}
        \\
         & \hspace{-1ex}{}\times
        \begin{pmatrix}
            1  & 1 & K \\
            -p & p & 0
        \end{pmatrix}
        \begin{pmatrix}
            1     & 1   & K        \\
            -\mu' & \mu & \mu'-\mu
        \end{pmatrix}
        \\
         & \hspace{-1ex}{}\times
        \begin{pmatrix}
            l & l' & L \\
            0 & 0  & 0
        \end{pmatrix}
        \begin{pmatrix}
            l  & l' & L        \\
            -m & m' & \mu'-\mu
        \end{pmatrix}
        \\
         & \hspace{-1ex}{}\times
        A_{KL}^{\mu-\mu'}[W].
    \end{aligned}
\end{equation}
where the parentheses denote Wigner $3j$ symbols\cite{zare_angular_1988},
and $A_{KL}^{q}[W]$ contains the remaining dependence on the orientational
distribution.
For convenience, the laboratory and molecular coordinate systems are chosen
such that their $z$ axes are parallel to $\hat e$ and $\hat a$,
respectively. The channel amplitudes $\tilde D_{lm\mu}$ are transformed to
this molecular alignment frame. With this convention, $\vartheta$ is the
polar Euler angle between the molecular and laboratory $z$ axes, and the
orientation factor takes the form
\begin{equation}
    \label{eq:axial_orientation_factor}
    A_{KL}^{q}[W]
    =
    \frac{1}{2}
    \int_{0}^{\pi}
    W(\vartheta)\,
    d^{K}_{0q}(\vartheta)\,
    d^{L}_{0q}(\vartheta)\,
    \sin\vartheta\,\du\vartheta\,,
\end{equation}
where $d^{J}_{mn}$ denotes a reduced Wigner rotation matrix\cite{zare_angular_1988}.

For an isotropic ensemble, the orientational weight is constant. Under the
normalization convention
\begin{equation*}
    \frac{1}{2} \int_0^\pi W(\vartheta)\sin\vartheta\,\du\vartheta = 1\,,
\end{equation*}
one has
\begin{equation}
    \label{eq:isotropic_orientation_weight}
    W_{\mathrm{iso}}(\vartheta)=1\,.
\end{equation}
Consequently, the particular choice of the molecular reference direction
$\hat a$ is immaterial for the isotropic ensemble. For a linearly polarized
probe pulse ($p=0$), only $b_0$ and $b_2$ are nonzero\cite{cooper_angular_1968,cooper_erratum_1968}.
For a circularly
polarized probe pulse ($p=\pm1$), an additional $b_1$ coefficient can be
nonzero for a chiral sample\cite{ritchie_theory_1976}.

After one-photon excitation by a linearly polarized pump pulse, the excitation
probability is proportional to the squared projection of the transition-dipole
direction at the geometry of excitation onto the pump-polarization direction.
With the preceding alignment-frame convention, these directions define the
molecular and laboratory $z$ axes, respectively. This produces the axially
symmetric orientational
distribution\cite{reid_extracting_2000,hockett_general_2015}
\begin{equation}
    \label{eq:prealigned_orientation_weight}
    W_{\mathrm{align}}(\vartheta)
    =
    3\cos^2\vartheta\,.
\end{equation}
This model neglects rotational dynamics during the pump--probe delay.
The resulting static-alignment approximation is appropriate at ultrashort delays,
before rotational motion appreciably changes the alignment distribution.
To retain the axial symmetry of the
combined pump--probe arrangement, a linearly polarized probe pulse ($p=0$) must be
polarized parallel to the pump pulse. For a circularly polarized probe pulse ($p=\pm 1$),
its propagation direction must be parallel to the pump-polarization
direction.
Only for these geometries is the pre-aligned ensemble described by the
axially symmetric form used in
Eqs.~\eqref{eq:axial_legendre_coefficients}--\eqref{eq:axial_orientation_factor}\cite{hockett_general_2015}.

Since $W_{\mathrm{align}}(\vartheta)=1+2P_2(\cos\vartheta)$, the alignment
distribution contains angular components of ranks zero and two. Coupling the
rank-two alignment component to the rank-two angular dependence of linearly
polarized one-photon ionization permits Legendre coefficients through $b_4$.
For the linearly polarized pump--probe geometry considered here, the odd
coefficients $b_1$ and $b_3$ vanish by
symmetry\cite{reid_extracting_2000,hockett_general_2015}.

In practical calculations, the continuum partial-wave expansion is truncated at
a maximum angular momentum $l_{\mathrm{max}}$. Thus, $0\leq l,l'\leq
    l_{\mathrm{max}}$, with $-l\leq m\leq l$ and $-l'\leq m'\leq l'$, in
Eqs.~\eqref{eq:partial_wave_dipole_amplitudes},
\eqref{eq:partial_wave_overlap}, \eqref{eq:axial_legendre_coefficients}, and
\eqref{eq:axial_rotational_averaging_coefficients}.

The partial-wave dipole amplitudes $D_{lm\mu}$ in
Eq.~\eqref{eq:partial_wave_dipole_amplitudes} are first calculated for each
ionization channel on a radial--Lebedev product grid. The radial integrals are
evaluated by Gauss--Legendre quadrature, whereas the angular integrals over the
unit sphere are evaluated by Lebedev quadrature. When explicit continuum
orthogonalization is used, the same quadrature grid is used to evaluate the
continuum--Dyson overlaps $S_{lm}$ and the Dyson-orbital dipole moments $M_\mu$
(\emph{cf.} Eqs.~\eqref{eq:partial_wave_overlap} and
\eqref{eq:dyson_spherical_dipole_moment}). For a pre-aligned ensemble, the
resulting channel amplitudes are transformed to the molecular alignment frame
before the contraction in Eq.~\eqref{eq:axial_legendre_coefficients} is
performed. The orientation factor $A_{KL}^{q}[W]$ in
Eq.~\eqref{eq:axial_orientation_factor} is evaluated by Gauss--Legendre
quadrature.

In the present work, $l_{\mathrm{max}}=10$ was used to obtain converged
results. Photoelectron spectra at the ground-state equilibrium geometry were
computed using squared Dyson norms alongside explicit plane-wave and
Coulomb-wave continuum approximations. The plane-wave continuum was expanded
about the Dyson centroid, whereas the Coulomb-wave continuum was expanded about
either the Dyson centroid or the charge centre. For the time-resolved spectra,
the ionization intensities along the nonadiabatic trajectories were modeled
using either squared Dyson norms or photoionization matrix elements evaluated
with the plane-wave and Coulomb-wave continuum approximations described above.
The continuum-wave results were averaged isotropically while the molecule
remained in the ground state and over the pre-aligned ensemble after
excitation, using the transition-dipole direction of the populated excited
state at the geometry of excitation as the alignment axis. The static-alignment
approximation is justified over the \SI{215}{\fs} simulation window, which is
much shorter than the rotational-revival timescales of several tens of
picoseconds or longer inferred from measured rotational constants of
furan.\cite{rukiye_ozer_masscorrelated_2024} The implementation is available as
the \texttt{PESKit} package at \url{https://github.com/mitric-lab/peskit}.

\section{Results and discussion}

\subsection{Electronic properties of furan}

The relevant geometries and electronic states involved in the nonadiabatic
dynamics of furan will be characterized in this section. The equilibrium
geometry of furan in its electronic ground state of furan was optimized at the
SA5-RMS-CASPT2(10, 10)/aug-cc-pVDZ lavel. This geometry is essentially
identical to that optimized at the CCSD(T)/cc-pVTZ
level\cite{purvis_full_1982,dunning_gaussian_1989,neese_software_2022}, with an
RMSD of \SI{0.016}{\angstrom}. The vertical excitation energies calculated at
this geometry are compared with various computational and experimental results
as shown in Table~\ref{tab:benchmark}. The SA5-RMS-CASPT2(10,10)/aug-cc-pVDZ
and CCSD(T)/cc-pVTZ equilibrium geometries, together with the corresponding
vertical excitation energies, are provided in
Sec.~\ref{sec:si-equilibrium-geometries} of the Supporting Information.

\begin{table*}[t]
    \caption{Theoretical and experimental vertical excitation energy in eV for the
        first six electronically excited singlet states of furan.}
    \label{tab:benchmark}
    \begin{tabular}{lcccccc}
        \hline
        state                            & this work & CASPT2\textsuperscript{\emph{a}} & NEVPT2\textsuperscript{\emph{b}} & MRMP\textsuperscript{\emph{c}} & CC3\textsuperscript{\emph{d}} & exp.\textsuperscript{\emph{e}} \\
        \hline
        $1\ ^1A_2 (\pi \to \mathrm{3s})$ & 5.83      & 5.92                             & 6.13                             & 5.84                           & 6.08                          & 5.91                           \\
        $1\ ^1B_2 (\pi \to \pi^*)$       & 5.98      & 6.04                             & 6.41                             & 5.95                           & 6.34                          & 6.04                           \\
        $2\ ^1A_1 (\pi \to \pi^*)$       & 6.04      & 6.16                             & 6.62                             & 6.16                           & 6.58                          & --                             \\
        \hline
    \end{tabular}

    \textsuperscript{\emph{a}}~From Ref.~\citenum{serrano-andres_theoretical_1993};
    \textsuperscript{\emph{b}}~PC-QDNEVPT2 results for the (0805) and (4322) active spaces from Ref.~\citenum{pastore_application_2006};
    \textsuperscript{\emph{c}}~From Ref.~\citenum{christiansen_electronic_1998};
    \textsuperscript{\emph{d}}~From Ref.~\citenum{loos_mountaineering_2020};
    \textsuperscript{\emph{e}}~Summarized by Wan \emph{et al.}\cite{wan_electronic_2000} from Refs.~\citenum{pickett_vibrational_1940,price_absorption_1941,pickett_vacuum_1951,derrick_photoelectron-spectroscopical_1971,robin_higher_1975,flicker_electron_1976,flicker_trlplet_1976,cooper_resonantly_1980,roebber_electronic_1980,palmer_electronic_1995}.
\end{table*}

All high-level calculations show good agreement with experiment for the
excitation energies of the $(\pi \to \mathrm{3s})$ Rydberg state and the first
$(\pi \to \pi^*)$ state. The $2\,^1A_1$ state is more challenging, however,
because it exhibits substantial non-single-excitation character. In our
RMS-CASPT2 calculation, singly excited configurations account for only about
\SI{74}{\percent} of this state, indicating a significant admixture of doubly
and possibly higher excited configurations, consistent with previous
theoretical
work\cite{christiansen_integral-direct_1998,gromov_theoretical_2003}.
Consequently, methods limited to singles tend to provide a less balanced
description, whereas approaches including doubles yield more reliable
excitation energies. The excitation energy obtained in our work is consistent
with recent high-level benchmarks, including
EOM-CCSDT\cite{loos_mountaineering_2020} and the theoretical best estimate from
the \textsc{Quest} database\cite{loos_quest_2025}. The 3p Rydberg states are
not considered here. The reasons for their exclusion are discussed in
Sec.~\ref{sec:si-3p-rydberg-exclusion} of the Supporting Information.

\begin{figure}[!htbp]
    \centering
    \includegraphics[scale=1.0]{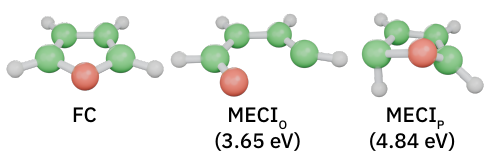}
    \caption{Optimized ground state geometry (FC),
        the ring-opening MECI (MECI\textsubscript{o}), and the
        ring-puckering MECI (MECI\textsubscript{p}) of furan.}
    \label{fig:furan_geoms}
\end{figure}

\begin{figure}[!htbp]
    \centering
    \includegraphics[scale=1.0]{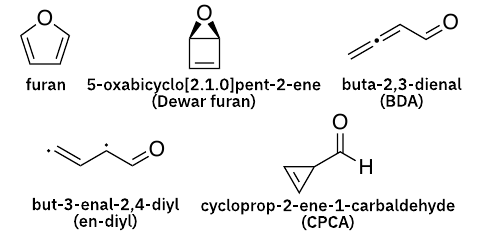}
    \caption{Chemical structures of the five possible photoproducts
    considered in this work:
    furan,
    5-oxabicyclo[2.1.0]pent-2-ene (Dewar furan),
    buta-2,3-dienal (BDA),
    but-3-enal-2,4-diyl (en-diyl),
    and cycloprop-2-ene-1-carbaldehyde (CPCA).}
    \label{fig:furan_products}
\end{figure}

\begin{figure}[!htbp]
    \centering
    \includegraphics[scale=1.0]{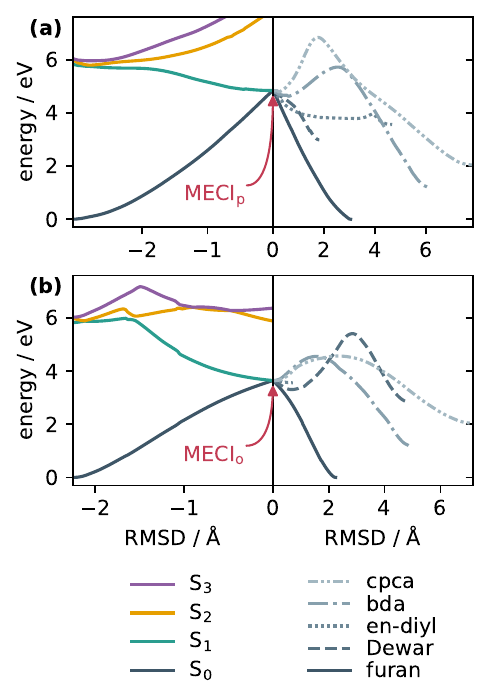}
    \caption{Interpolated energy scans of furan connecting the
        optimized ground state geometry (FC) to five possible photoproducts
        \emph{via} (a) the ring-puckering MECI and (b) the ring-opening MECI.
        The left panels show the energies of the four lowest singlet states
        along the interpolation from FC to the respective MECI,
        and right panels show the ground-state energy along the
        interpolation from the MECI to the product structures shown
        in Fig.~\ref{fig:furan_products}.}
    \label{fig:furan_meci_product_scan}
\end{figure}

It is well known that two S\textsubscript{0}/S\textsubscript{1} minimum-energy
conical intersections (MECIs), namely the ring-puckering one
(MECI\textsubscript{p}) and the ring-opening one (MECI\textsubscript{o}), play
an important role in the relaxation dynamics of
furan\cite{wilsey_role_1996,stenrup_computational_2010_utf,lindner_multistate_2018,adachi_probing_2018}.
We optimized the geometries of these two MECIs using the projected constrained
optimization
method\cite{anglada_reduced-restricted-quasi-newton-raphson_1997,de_vico_new_2005},
which are shown, along with the optimized ground state geometry, in
Fig.~\ref{fig:furan_geoms}. Their Cartesian coordinates are provided in
Sec.~\ref{sec:si-meci-geometries} of the Supporting Information. In addition,
we considered five possible photoproducts of furan, whose chemical structures
are shown in Fig.~\ref{fig:furan_products}. To assess the balance of the active
space not only around the Franck--Condon region and the MECIs, but also along
reaction pathways connecting these structures to the products, we performed two
sets of interpolated\cite{zhu_geodesic_2019} scans: one connecting furan to the
product structures \emph{via} MECI\textsubscript{p}, and one \emph{via}
MECI\textsubscript{o}. The resulting energy curves are shown in
Fig.~\ref{fig:furan_meci_product_scan}. The smooth potential energy profiles
along both sets of scans indicate that the chosen active space and averaging
scheme provide a balanced description along these pathways.

\begin{figure}[!htbp]
    \centering
    \includegraphics{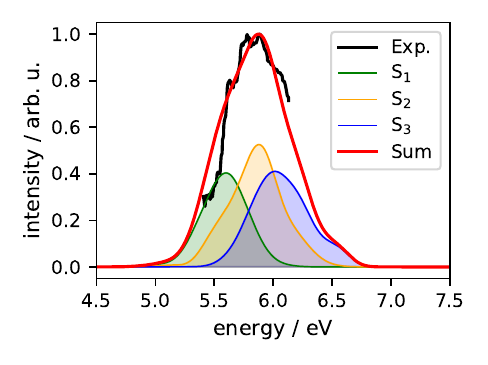}
    \caption{Comparison of the simulated ensemble spectrum
        and the experimental absorption spectrum of furan.
        The simulated spectrum is obtained using the SA5-RMS-CASPT2(10, 10)/aug-cc-pVDZ
        method with 250 geometries sampled from a harmonic canonical Wigner distribution
        at \SI{50}{\K}, and convolved with a Gaussian profile with
        $\sigma=\SI{0.1}{\eV}$.
        The contribution of each electronically excited state is shown
        with thin lines. The experimental spectrum is taken from
        Ref.~\citenum{roebber_electronic_1980}.}
    \label{fig:ensemble_uvvis}
\end{figure}

Although only the $\mathrm{S_0}\to \mathrm{S_2}$ transition is bright among the
first three singlet excited states at Franck-Condon geometry,
S\textsubscript{1} and S\textsubscript{3} can gain intensities through vibronic
interaction with
S\textsubscript{2}\cite{gromov_theoretical_2003,koppel_multi-modemulti-state_2004,gromov_theoretical_2004},
thus altering the spectral shape. Therefore, an ensemble spectrum is simulated
using the vertical excitation energies and oscillator strengths of the 250
initial geometries generated for the dynamics simulations, and convolved with a
Gaussian profile of $\sigma = \SI{0.1}{\eV}$. A comparison with jet-cooled
absorption spectrum of furan recorded by Roebber \emph{et
    al.}\cite{roebber_electronic_1980} is shown in Fig.~\ref{fig:ensemble_uvvis}.

The simulated ensemble spectrum is in good agreement with the experimental one.
Its band shape is formed by contributions from transitions to the adiabatic
S\textsubscript{1}, S\textsubscript{2}, and S\textsubscript{3} states across
the geometries of the initial ensemble, with substantial geometry-dependent
redistribution of oscillator strength among these states. At the equilibrium
geometry, in contrast, the stick spectrum is dominated by the bright
S\textsubscript{2} transition, and an appropriate broadening of this stick
spectrum already reproduces the experimental band reasonably well. However, the
resulting spectrum remains shifted, being centered at about \SI{6.0 }{eV}, and
because the experimental band is itself close to Gaussian, line-shape
comparisons alone are of limited diagnostic value. This clearly demonstrates
that promoting a trajectory to any single adiabatic state does not faithfully
represent the excitation process or the subsequent nonadiabatic relaxation. We
therefore employ the FISH method with explicit laser field excitation to
perform the dynamics simulation.

For the characterization of the cationic states of furan, vertical ionization
energies were computed using SA4-RMS-CASPT2(9,10), SA8-RMS-CASPT2(9,10), and
SA10-RMS-CASPT2(13,12). The SA4 and SA8 calculations were referenced to the
neutral ground state described at the SA5-RMS-CASPT2(10,10) level, whereas SA10
was referenced to SA5-RMS-CASPT2(14,12). The active orbitals were chosen in
close analogy to those used for the neutral excited-state calculations. For
consistency, the 3s Rydberg orbital was retained in the cationic active space
by freezing it during the CASSCF optimization, as it would otherwise rotate
out. In comparison to SA4 and SA8, the enlarged SA10-RMS-CASPT2(13,12) active
space includes two additional occupied \textsigma orbitals in order to probe
additional \textsigma-ionization channels. The corresponding natural orbitals
are shown in Fig.~\ref{fig:si-sa10-natural-orbitals} of the Supporting
Information.

\begin{figure}[!htbp]
    \centering
    \includegraphics{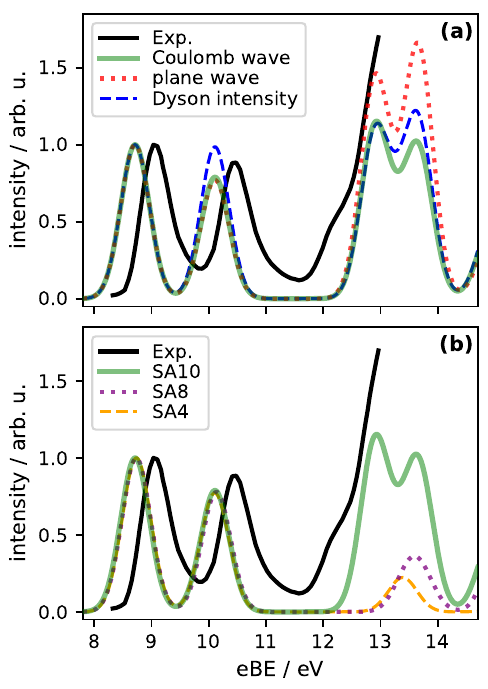}
    \caption{Comparison of the experimental probe-only photoelectron spectrum (PES) of furan
        reproduced from Ref.~\citenum{uenishi_signatures_2024} with simulated PES obtained
        using (a) squared Dyson norms, plane-wave matrix elements, and Dyson-centered
        Coulomb-wave matrix elements based on SA10-RMS-CASPT2(13, 12), and
        (b) Dyson-centered Coulomb-wave matrix elements based on the
        SA10-RMS-CASPT2(13, 12), SA8-RMS-CASPT2(9, 10), and
        SA4-RMS-CASPT2(9, 10) methods. The charge-centered Coulomb-wave spectrum is not
        shown in panel (a), because it is visually indistinguishable from the
        Dyson-centered Coulomb-wave spectrum on this scale. Gaussian broadening with
        $\sigma=\SI{0.25}{\eV}$ was applied.}
    \label{fig:pes_comparison}
\end{figure}

Before comparing different cationic-state manifolds, we assess the
approximation used for the relative photoionization intensities.
Figure~\ref{fig:pes_comparison}(a) compares the experimental probe-only
photoelectron spectrum\cite{uenishi_signatures_2024} with simulated spectra
obtained from squared Dyson norms, plane-wave matrix elements, and
Dyson-centered Coulomb-wave matrix elements, all normalized to the first band.
The corresponding charge-centered Coulomb-wave spectrum is not shown separately
because it is visually indistinguishable from the Dyson-centered Coulomb-wave
result on the scale of the figure. Its SA8 spectrum is provided in
Fig.~\ref{fig:si-ensemble-pes} of the Supporting Information. The three
theoretical spectra were computed using the same underlying SA10-RMS-CASPT2
description. A Gaussian broadening with $\sigma=\SI{0.25}{\eV}$ was used.

All three intensity models reproduce the first two bands with similar accuracy,
although they differ in the relative height of the second peak. In particular,
the plane-wave and Coulomb-wave spectra are close in this region, consistent
with the high-kinetic-energy limit in which the Coulomb-wave description
approaches the plane-wave result. More pronounced differences occur in the
higher-binding-energy region around \SIrange[range-phrase=--]{13}{14}{\eV},
where the relative intensities become more sensitive to the continuum
treatment. In this region, the plane-wave spectrum gives the closest visual
agreement with the experimental peak pattern. Nevertheless, the Coulomb-wave
intensities remain reasonable, and the Coulomb-wave continuum has the correct
long-range asymptotic form for photoionization into a positively charged
residual ion. For this reason, Coulomb-wave matrix elements are used for the
remaining photoelectron spectra discussed below.

Figure~\ref{fig:pes_comparison}(b) compares the experimental probe-only
photoelectron spectrum with simulated spectra obtained from Dyson-centered
Coulomb-wave matrix elements for the SA4-, SA8-, and SA10-RMS-CASPT2
cationic-state manifolds. The peaks at \SI{9.1}{\eV} and \SI{10.5}{\eV}
essentially correspond to ionization from the $\pi_3$ and $\pi_2$ orbitals,
respectively (\emph{cf.} Fig.~\ref{fig:furan_nos}). These two bands are
reproduced well by all three cationic-state manifolds, apart from an overall
shift of about \SI{0.3}{\eV}, indicating that the low-binding-energy part of
the spectrum is already captured at the SA4 level.

The differences among the SA4, SA8, and SA10 spectra arise mainly at higher
binding energies. The SA4 manifold is clearly insufficient in this region,
because it does not include enough cationic states to describe the higher-lying
ionization channels. Although SA8 improves upon SA4, it still strongly
underestimates the intensity in the \SIrange[range-phrase=--]{12}{14}{\eV}
region when Coulomb-wave matrix elements are used. Based on Koopmans-type
estimates,\cite{baerends_density_2018,wan_electronic_2000} additional
\textsigma-ionization channels are expected to contribute in this energy range,
and these channels are only incompletely represented at the SA8 level. The SA10
manifold captures substantially more of this high-binding-energy structure, but
requires a substantially higher computational effort. Since the
low-binding-energy $\pi$-ionization bands are already well described at the SA8
level, and since SA8 is considerably more affordable than SA10, the SA8
manifold was adopted for the calculations along the nonadiabatic trajectories.

The photoelectron spectra obtained from the 250 Wigner-sampled geometries at
the SA8-RMS-CASPT2(9,10) level differ only minimally from the corresponding
single-geometry spectra, irrespective of whether the relative intensities are
estimated from squared Dyson norms, plane-wave matrix elements, or Coulomb-wave
matrix elements with either choice of origin, as shown in
Fig.~\ref{fig:si-ensemble-pes} of the Supporting Information.

\begin{figure}[!htbp]
    \centering
    \includegraphics{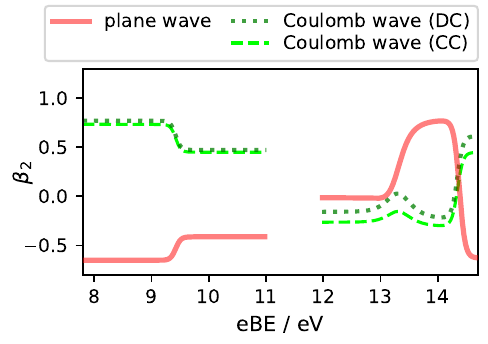}
    \caption{Photoelectron anisotropy parameter $\beta_2$ obtained from
        plane-wave matrix elements and from Coulomb-wave matrix elements using either
        the Dyson centroid (DC) or the charge center (CC) as the Coulomb-wave origin.
        All curves were computed using the SA10-RMS-CASPT2(13, 12) electronic-structure
        description. The discrete $\beta_2$ values were broadened with Gaussians of
        $\sigma=\SI{0.25}{\eV}$ by taking the ratio of the broadened anisotropy-weighted
        intensity to the broadened intensity. Values are shown only where the broadened
        intensity exceeds $10^{-3}$ of its maximum.}
    \label{fig:beta2_comparison}
\end{figure}

The effect of the continuum approximation is more pronounced in the
photoelectron anisotropy parameter than in the angle-integrated PES.
Figure~\ref{fig:beta2_comparison} compares $\beta_2$ obtained from plane-wave
matrix elements and from Coulomb-wave matrix elements using either the Dyson
centroid or the charge center as the Coulomb-wave origin. Although the
Dyson-centered and charge-centered Coulomb waves give nearly indistinguishable
angle-integrated spectra, their $\beta_2$ values show visible differences,
especially in the higher-binding-energy region. The plane-wave and Coulomb-wave
results also differ substantially, indicating that the apparent similarity of
some angle-integrated spectra does not imply an equally similar angular
distribution. The anisotropy parameter therefore provides a more sensitive
probe of the continuum approximation. The corresponding single-point and
250-geometry Wigner-ensemble $\beta_2$ spectra at the SA8-RMS-CASPT2(9, 10)
level are provided in Fig.~\ref{fig:si-ensemble-beta2} of the Supporting
Information.

\subsection{Nonadiabatic dynamics simulation}
\begin{figure}[!htbp]
    \centering
    \includegraphics{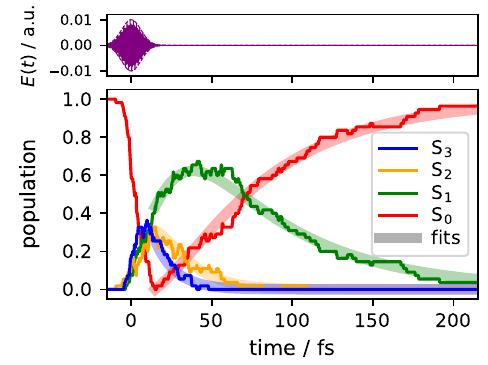}
    \caption{Electric field $E(t)$ of the excitation pulse in atomic units, its
        Gaussian envelope (upper panel), and time evolution of the averaged
        populations of the first four singlet states of furan during the FISH
        dynamics simulation (lower panel). The thick, semi-transparent lines
        are the fits to the population decay using a sequential decay model.}
    \label{fig:population}
\end{figure}

For the 55 trajectories excited by the laser pulse, the time evolution of the
averaged electronic state populations obtained from FISH dynamics is shown in
Fig.~\ref{fig:population}.

The excitation process completes after approximately \SI{30}{\fs}, at which
point the populations of states S\textsubscript{1}, S\textsubscript{2}, and
S\textsubscript{3} are approximately \SI{42}{\percent}, \SI{33}{\percent}, and
\SI{25}{\percent}, respectively. After the pulse, the S\textsubscript{3} and
S\textsubscript{2} populations decay rapidly, followed by a slower relaxation
of S\textsubscript{1} to the ground state. A sequential fit to the post-pulse
populations yields time constants of \SI{13}{\fs}, \SI{13}{\fs}, and
\SI{64}{\fs} for the processes $\mathrm{S_3}\to\mathrm{S_2}$,
$\mathrm{S_2}\to\mathrm{S_1}$, and $\mathrm{S_1}\to\mathrm{S_0}$, respectively.
A visualization of the fit is shown as thick lines in
Fig.~\ref{fig:population}. The fitted constants are of the same overall order
as those reported previously for furan in nonadiabatic simulations
\cite{fuji_ultrafast_2010,oesterling_substituent_2017_utf,filatov_signatures_2021}.
At the end of the simulation time (\SI{215}{\fs}), 53 of the 55 trajectories
are relaxed to the electronic ground state, while the other two stay in the
S\textsubscript{1} state.

Classification of the stuctures at which the
S\textsubscript{1}$\to$S\textsubscript{0} transition happens shows that most
trajectories relax through geometries closer to MECI\textsubscript{P} than to
MECI\textsubscript{O}. This preference is consistent with TRPES studies
interpreted with TDDFT dynamics \cite{fuji_ultrafast_2010,adachi_probing_2018},
but contrasts with attosecond X-ray absorption \cite{severino_attosecond_2024}
and Coulomb-explosion imaging \cite{wang_time-resolved_2023}, both aided by
TDDFT dynamics, as well as with CASSCF surface-hopping simulations
\cite{hua_monitoring_2016,oesterling_substituent_2017}, which emphasize ring
opening. However, when the trajectories are classified instead by the geometry
at which the S\textsubscript{1}--S\textsubscript{0} gap becomes minimal, the
ring-opening contribution increases substantially. This shows that a
significant fraction of trajectories develop ring-opened structures after
passing through the dominant ring-puckering deactivation funnel, thereby
reconciling a predominantly puckering-mediated internal conversion with
experimental signatures of ring opening. Consistent with this interpretation,
the interpolated S\textsubscript{0} scan from MECI\textsubscript{P} to en-diyl
does not exhibit an additional barrier
(Fig.~\ref{fig:furan_meci_product_scan}), indicating that trajectories passing
through the puckering funnel can subsequently access an open-chain structure.
Aggregate statistics are given in Table~\ref{tab:si-trajectory-classification},
and trajectory-resolved assignments and times are given in
Table~\ref{tab:si-trajectory-details} of the Supporting Information.

Because a low barrier has been reported on the pathway to
MECI\textsubscript{O}\cite{gavrilov_deactivation_2008,gromov_theoretical_2010,stenrup_computational_2010_utf,filatov_signatures_2021},
and is found in our calculation (Fig.~\ref{fig:furan_meci_product_scan}),
whereas the puckering route is formally barrierless, it is often argued that
the formation of MECI\textsubscript{O} is kinetically disfavored. However, our
RMS-CASPT2 scan places this barrier at \SI{5.99}{\eV} above the
S\textsubscript{0} Franck--Condon point, \emph{i.e.} essentially at the
S\textsubscript{2} excitation energy. Given the spectral width of the
excitation band (\emph{cf.} Fig.~\ref{fig:ensemble_uvvis}) and the virtually
identical hop times for both MECIs, these results suggest that access to
MECI\textsubscript{O} is not significantly kinetically impeded.

\begin{figure}[!htbp]
    \centering
    \includegraphics[width=\linewidth]{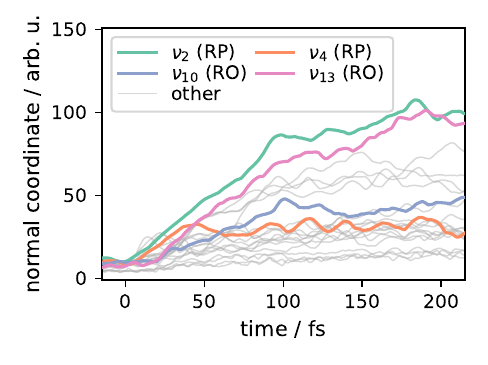}
    \caption{Projection of the trajectory displacements onto the normal modes of furan
        at the Franck-Condon point. The normal modes are labeled according to their
        frequency at the Franck-Condon point, and the displacements are averaged
        over all 55 trajectories.
        The ring-puckering modes $\nu_2$ and $\nu_4$ as well as the
        ring-opening modes $\nu_5$ and $\nu_6$ are highlighted.}
    \label{fig:traj_nm_projection}
\end{figure}

Instead, the shape of the excited-state potential-energy surfaces appears to
control the initial preference. Harmonic analysis at the Franck--Condon point
predicts that the ring-opening coordinate should be most unstable on
S\textsubscript{2} (imaginary frequency \complexqty{2499i}{\per\cm}), whereas
S\textsubscript{1} and S\textsubscript{3} are destabilized primarily along
ring-puckering modes (\complexqty{277i}{\per\cm} on S\textsubscript{1},
\complexqty{245i}{\per\cm} and \complexqty{700i}{\per\cm} on
S\textsubscript{3}). Surprisingly, trajectories initiated on S\textsubscript{2}
show the lowest MECI\textsubscript{O} yield. This suggests that dynamics
studies that launch furan exclusively on the S\textsubscript{2} surface may
overestimate the contribution of the ring-puckering pathway. Inspection of the
trajectories reveals that population launched on S\textsubscript{2} hops to
S\textsubscript{1} within \emph{ca.} \SI{13}{\fs}, well before significant
nuclear amplitude builds up along the ring-opening coordinate, so the system is
diverted onto the S\textsubscript{1} surface, where puckering dominates. Thus,
nonadiabatic transfer timescales and anharmonic mode coupling outweigh the
simple harmonic picture in steering the early dynamics. To visualize the
vibrational energy transfer, the trajectory displacements are projected onto
the normal modes of the Franck-Condon point. The average over all trajectories
is shown in Fig.~\ref{fig:traj_nm_projection}. It can be seen that the
vibrational energy gained upon excitation is immediately transferred to the
ring-puckering modes ($\nu_2$ and $\nu_4$), whereas the ring-opening modes are
energized with a delay of \emph{ca.} \SI{20}{\fs}.

Taken together, the data indicate that ring puckering provides the dominant
doorway for internal conversion at \SI{200}{\nm}, while rapid vibrational
energy redistribution on S\textsubscript{0} channels a substantial fraction of
population into the ring-opening valley, reconciling the apparent discrepancies
among previous experimental and theoretical studies.

\begin{figure*}[t]
    \centering
    \includegraphics[width=\linewidth]{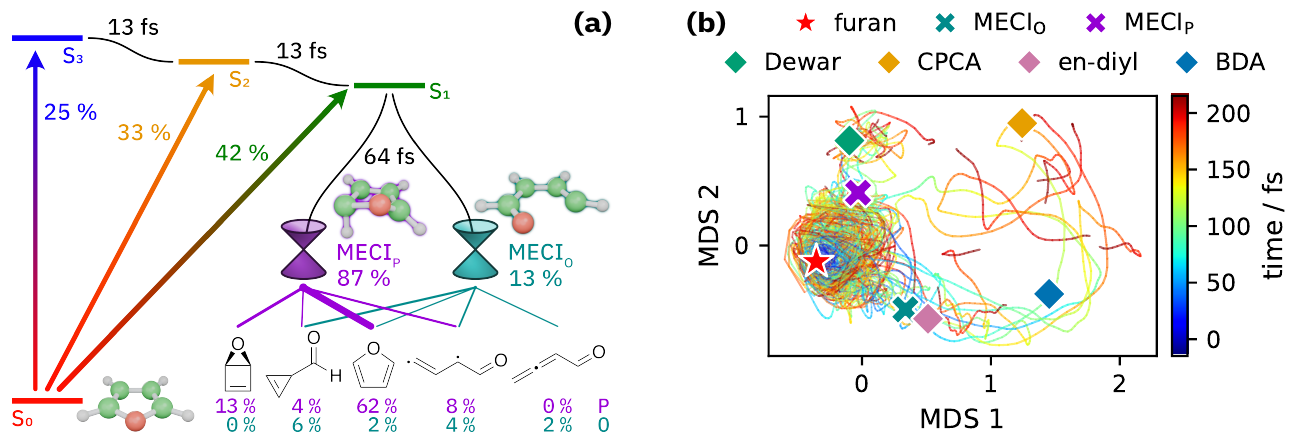}
    \caption{%
        Overview of the excited-state relaxation and product formation observed in
        the trajectory ensemble. (a) Schematic representation of the relaxation
        pathways. Widths of the colored population-flow paths encode trajectory
        counts qualitatively. Percentages next to the excitation paths give the
        distribution among the initially populated electronic states for the 55
        excited trajectories, and the numbers above the relaxation paths give the
        time constants obtained from the sequential fit to the population dynamics.
        Of these trajectories, 53 returned to S\textsubscript{0} within the
        simulation time. The percentages next to MECI\textsubscript{P} and
        MECI\textsubscript{O} give the corresponding fractions of these 53
        trajectories. The upper purple (P) and lower teal (O) percentages beneath
        each product give the fractions of the same 53 trajectories assigned to
        that product, classified according to the geometry at the
        S\textsubscript{1}/S\textsubscript{0} hop. Percentages are rounded to the
        nearest integer and may therefore not sum to \SI{100}{\percent}.
        (b) Two-dimensional multidimensional-scaling (MDS) embedding of the
        geometries sampled by the 55 trajectories and of the optimized reference
        structures. Lines trace individual trajectories and are colored according
        to simulation time. Symbols mark furan, the two MECIs, and the product
        reference structures. To enhance the visibility of trajectory segments
        in sparsely populated regions, their opacity and width are weighted
        inversely by the local density in the embedding.
    }
    \label{fig:relaxation_scheme}
\end{figure*}

At the end of the simulations, structural classification of all 55 excited
trajectories gives five classes: furan (\SI{65}{\percent}),
5-oxabicyclo[2.1.0]pent-2-ene (Dewar furan; \SI{13}{\percent}),
but-3-enal-2,4-diyl (en-diyl; \SI{11}{\percent}), buta-2,3-dienal (BDA;
\SI{2}{\percent}), and cycloprop-2-ene-1-carbaldehyde (CPCA; \SI{9}{\percent}),
all of which except for the biradical species could be experimentally confirmed
\emph{via} matrix isolation\cite{rendall_dewar_1986}. While Dewar furan is
formed \emph{via} ring-puckering motion, the products en-diyl, BDA, and CPCA
result from ring-opening motion. Due to the large amount of vibrational energy
stored in the ring-puckering mode, the discrimination between hot furan and
Dewar furan is difficult, as mentioned by Adachi \emph{et
    al.}\cite{adachi_probing_2018}. However, due to the aforementioned vibrational
energy redistribution, the photoproducts cannot be directly correlated to their
traversed MECIs. For instance, trajectories hopping to S\textsubscript{0} at
MECI\textsubscript{P}-like geometries can subsequently form the ring-opened
en-diyl, whereas trajectories hopping at MECI\textsubscript{O}-like geometries
can return to furan or form CPCA. A detailed breakdown of the relationships
between traversed MECIs and final photoproducts is provided in
Table~\ref{tab:si-trajectory-details} of the Supporting Information. The
overall excited-state relaxation and product-branching mechanism is summarized
in Fig.~\ref{fig:relaxation_scheme}, schematically in panel (a) and as a
two-dimensional MDS embedding\cite{borg2005modern} of the sampled molecular
geometries in panel (b). Details of the structural distance measure and MDS
procedure underlying Fig.~\ref{fig:relaxation_scheme}(b), together with
comparisons to alternative dimensionality-reduction methods, are provided in
Sec.~\ref{sec:si-trajectory-embedding} of the Supporting Information.

Experiments show that the majority of the excited molecules relax to hot furan
in \SIrange{150}{260}{\fs}\cite{adachi_probing_2018}, while the others form
Dewar furan and CPCA. This interval overlaps the \SI{215}{\fs} span of our
RMS-CASPT2 trajectories, allowing a direct product comparison. The simulations
likewise predict hot furan as the main species, with Dewar furan and CPCA as
less-abundant products. In contrast, the en-diyl and BDA channels were not
found in the TRPES study probably due to the low quantum yield of BDA
(\emph{ca.} \SI{2}{\percent}) and, in the case of en-diyl, its rapid secondary
rearrangement. Independent evidence for ring-opened products comes from
attosecond X-ray absorption\cite{severino_attosecond_2024} and
Coulomb-explosion imaging\cite{wang_time-resolved_2023}, both of which report
transient signatures compatible with the open-ring species seen here.

Taken together, the five-product pattern obtained from our RMS-CASPT2
trajectories is consistent with existing ultrafast data. At longer delays, the
product distribution may evolve further. CPCA could form through rearrangement
of Dewar furan\cite{rendall_dewar_1986}, while the en-diyl biradical may
undergo secondary rearrangement or hydrogen transfer. These processes could
alter the product ratios beyond the simulated time window.

\subsection{Time-resolved photoelectron spectrum}

\begin{figure*}[t]
    \centering
    \includegraphics[width=\linewidth]{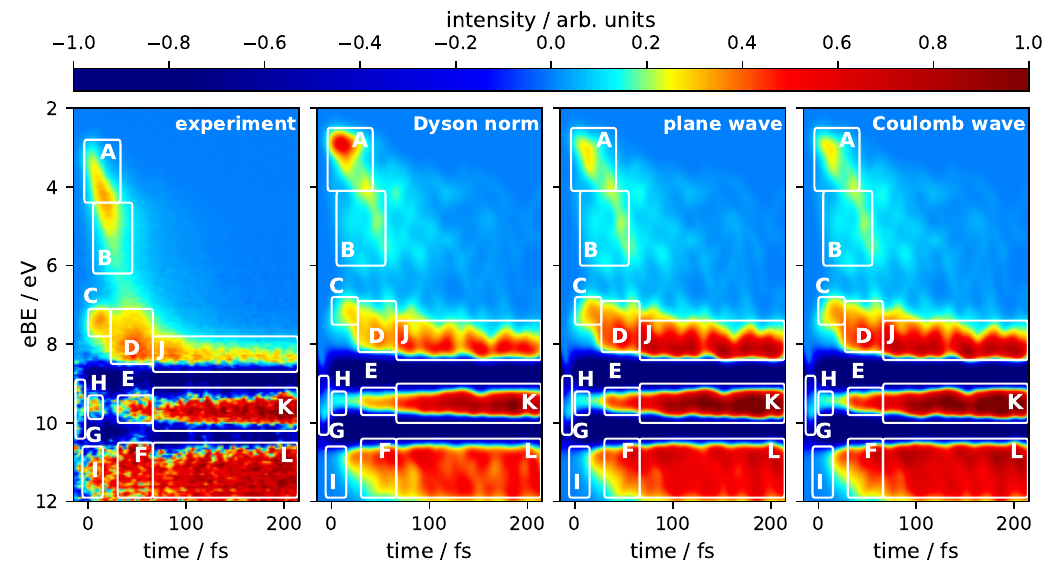}
    \caption{Comparison of the experimental time-resolved photoelectron spectrum
        (TRPES) of furan taken from Ref.~\citenum{uenishi_signatures_2024} with simulated
        TRPES obtained from squared Dyson norms, plane-wave photoionization matrix elements,
        and Coulomb-wave photoionization matrix elements. The electronic structure along
        the FISH dynamics trajectories was described using SA5-RMS-CASPT2(10, 10) and
        SA8-RMS-CASPT2(9, 10). The spectra were obtained by convolving the discrete
        ionization events with a two-dimensional Gaussian profile with
        $\sigma_t = \SI{4}{\fs}$ and $\sigma_{\mathrm{eBE}} = \SI{0.2}{\eV}$, followed by
        subtraction of the ensemble PES at the initial structures of all 250 trajectories.
        For the Coulomb-wave spectrum, the Coulomb-wave origin was placed at the Dyson
        centroid. The labeled regions (A--L) are discussed in the text.}
    \label{fig:trpes}
\end{figure*}

Along the 55 excited trajectories obtained from the FISH dynamics, ionization
energies and Dyson orbitals are calculated using SA5-RMS-CASPT2(10, 10) and
SA8-RMS-CASPT2(9, 10). The relative photoelectron intensity is modeled using
either squared Dyson norms or photoionization matrix elements evaluated with
plane-wave and Coulomb-wave continuum functions. The TRPES is obtained by
convolving the points using a two-dimensional Gaussian profile with $\sigma_t =
    \SI{4}{\fs}$ and $\sigma_{\mathrm{eBE}} = \SI{0.2}{\eV}$ and subtracting the
ensemble PES at the initial structures of all 250 trajectories, thereby
emulating the ground-state bleach present in the experimental transient PES.
The abbreviation eBE stands for electron binding energy. The resulting TRPES
are compared with the experimental TRPES from
Ref.~\citenum{uenishi_signatures_2024} in Fig.~\ref{fig:trpes}. The
experimental spectra have the pump-only and probe-only photoelectron
backgrounds subtracted. For the Coulomb-wave spectrum shown in
Fig.~\ref{fig:trpes}, the Coulomb-wave origin is placed at the Dyson centroid.
The corresponding charge-centered Coulomb-wave TRPES is not shown separately
because it is visually indistinguishable on the scale of the figure. The
isotropically averaged and pre-aligned TRPES for all three continuum-wave
treatments are compared in Fig.~\ref{fig:si-trpes-averaging-comparison} of the
Supporting Information.

The simulated TRPES obtained from squared Dyson norms shows good agreement with
the experimental one, except for the signal at an eBE of about \SI{3.0}{\eV}.
In the Dyson-norm spectrum, this signal bifurcates into two branches: one of
them remains at approximately the same eBE for about \SI{30}{\fs}, and the
other one shifts to \SI{4.0}{\eV} between \num{20} and \SI{40}{\fs}. In the
experiment, only the latter branch is observed. This discrepancy is strongly
reduced when the photoelectron intensity is calculated from approximate
continuum states: both the plane-wave and Coulomb-wave spectra show a weaker
low-eBE branch and therefore reproduce the shape of signal A more closely. This
suggests that, for signal A, the squared-Dyson-norm approximation overestimates
the relative intensity of the slowly shifting contribution. The contributions
from the trajectory sub-ensembles occupying S\textsubscript{1},
S\textsubscript{2}, and S\textsubscript{3} at time zero are shown in
Fig.~\ref{fig:si-excitation-separated-trpes} of the Supporting Information.

The TRPES exhibits a rich set of features. Twelve spectral regions are labeled
in the experimental spectrum, with the corresponding regions indicated in the
simulated spectra in Fig.~\ref{fig:trpes}. The time--eBE windows used for the
labeled regions (A--L) are given in Table~\ref{tab:si-trpes-signal-windows} of
the Supporting Information. From the trajectory-resolved ionization channels, a
tentative assignment of these signals is presented in the following. When
comparing with the assignments of Uenishi \emph{et al.}, the indices of the
neutral states S\textsubscript{x} are adapted, since they did not include the
3s Rydberg state in their indexing.

Signal A originates from ionization of excited furan to D\textsubscript{0}. The
branch that shifts rapidly in eBE is attributed to the $\text{S}_1 \to
    \text{D}_0$ and $\text{S}_2 \to \text{D}_0$ channels, while the slowly shifting
low-eBE branch is attributed to $\text{S}_3 \to \text{D}_0$. This latter
contribution is prominent in the Dyson-norm spectrum but is strongly reduced in
both continuum-based spectra. The improved agreement with experiment therefore
supports this assignment of the early-time ionization channels. The neighboring
signal B is mainly caused by the ionization $\text{S}_1 \to \text{D}_0$, with
minor contributions from $\text{S}_2 \to \text{D}_0$ and $\text{S}_{1/2} \to
    \text{D}_1$. Uenishi \emph{et al.} recognized and explained the sharp increase
in the rate of change of the vertical ionization energy (VBE), $\frac{\du
        (\mathrm{VBE})}{\du t}$, when transiting from signal A to B. From a potential
energy surface scan, they concluded that this rapid change is caused by the
character change of the S\textsubscript{2} state, resulting in a kink on the
potential energy curve along the reaction coordinates. Our simulation confirms
the kink (\emph{cf.} Fig.~\ref{fig:si-state-ionization-scans}). Furthermore,
the relaxation from S\textsubscript{3} and S\textsubscript{2} to
S\textsubscript{1} strengthens this increase.

Signal C at around \SI{7.5}{\eV} was observed for the first time by Uenishi
\emph{et al.}, and they assigned it to the $\text{S}_2 \to \text{D}_2$
transition. From our simulation, we can conclude that the final state is indeed
the D\textsubscript{2} state, but all the excited neutral states
(S\textsubscript{1}, S\textsubscript{2} and S\textsubscript{3}) contribute as
the initial state. As the molecular geometry becomes distorted, the ionization
channels to D\textsubscript{1} open (Fig.~\ref{fig:si-dyson-norm-scans}) and
contribute at \SIrange{30}{60}{\fs}, resulting in both $\text{S}_1 \to
    \text{D}_1$ and $\text{S}_1 \to \text{D}_2$ contributing to signal D. Due to
the rapid internal conversion to the S\textsubscript{1} state, ionization
channels from higher neutral states (S\textsubscript{2} and S\textsubscript{3})
are of little importance. Also, a few trajectories are already in
S\textsubscript{0} at this time delay, which results in a mixing with
ionization signals from S\textsubscript{0}. In particular, the channels
$\text{S}_0 \to \text{D}_{0/1}$ play a role in signal D, which increases with
time delay.

Uenishi \emph{et al.} recognized a kink in $\frac{\du (\mathrm{VBE})}{\du t}$
also at the transition from C to D. Our scans attribute this kink primarily to
changes in the character of the D\textsubscript{1} and D\textsubscript{2}
states, superimposed on the internal conversion to S\textsubscript{1}. The
$\text{S}_1 \to \text{D}_{2}$ ionization energy exhibits a kink at an RMSD of
approximately \SI{1.8}{\angstrom} toward MECI\textsubscript{P} and
\SI{1.3}{\angstrom} toward MECI\textsubscript{O}
(Fig.~\ref{fig:si-state-ionization-scans}). Because the evolution from
C$\,\to\,$D is distributed over the D\textsubscript{1} and D\textsubscript{2}
ionization channels and a broader range of geometrical distortions, the
corresponding change in $\frac{\du (\mathrm{VBE})}{\du t}$ is less pronounced
than for the A$\,\to\,$B transition.

In the simulation, regions E and F contain ionization contributions from the
S\textsubscript{1} and S\textsubscript{0} states, with the contribution from
S\textsubscript{0} increasing towards larger time delays. These contributions
appear in the same time interval as signal D. The major contributors in region
E are the $\text{S}_0 \to \text{D}_{0/1}$ and $\text{S}_1 \to \text{D}_{3/4}$
channels. Signal F mainly contains $\text{S}_0 \to \text{D}_{1/2}$ and
$\text{S}_1 \to \text{D}_{x}$ contributions with $x\geq 6$. Because some
\textsigma-orbitals are missing from the active space of (10, 10) (\emph{cf.}
Fig.~\ref{fig:pes_comparison}), the exact values of $x$ cannot be determined
from our simulation. This is probably also the reason that the simulated signal
F is weaker and has a slightly different shape than its experimental
counterpart.

The residual intensity in region G is comparable to negative-delay fluctuations
and is not established as a significant transient feature. In individual
simulated trajectories, the calculated branches at an eBE of \SI{9}{\eV} and
\SI{10}{\eV} correspond to the $\text{S}_0 \to \text{D}_0$ and $\text{S}_0 \to
    \text{D}_1$ transitions, respectively. The simulated contribution in region H
is attributed to the $\text{S}_1 \to \text{D}_{3/4}$ channels. Experimentally,
H and E are not clearly resolved as distinct components within the intensity
fluctuations. Signal I appears in the same early-time interval as region H.
Like signal F, it corresponds to $\text{S}_1 \to \text{D}_{x}$ with $x\geq 6$.
Again, due to the absence of some \textsigma-orbitals in the simulation, the
contributing channels cannot be precisely assigned, and the simulated signal I
is much weaker than the experimental one.

\begin{figure*}[!t]
    \centering
    \includegraphics[width=0.88\linewidth]{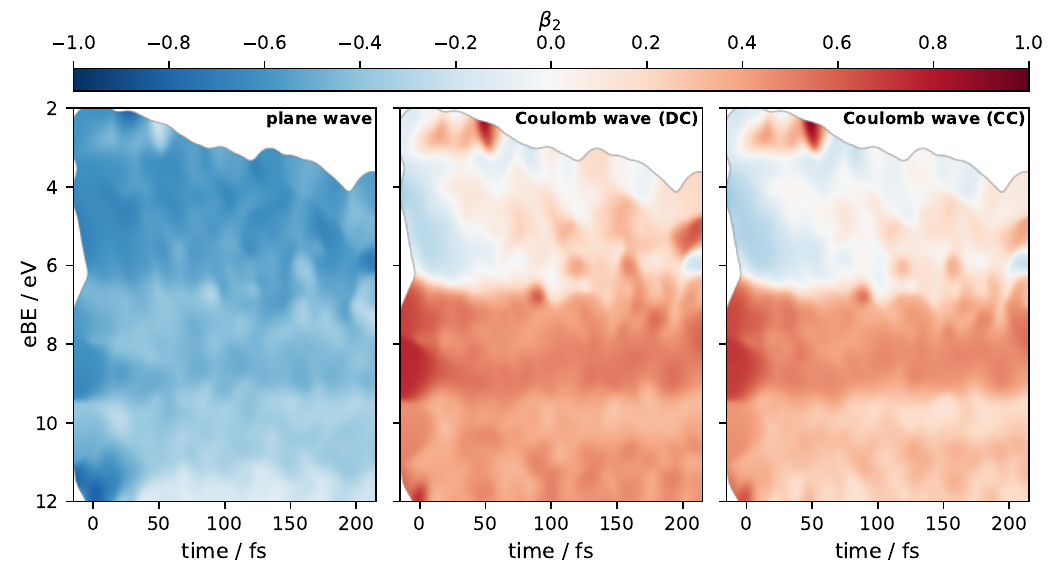}
    \par\vspace{-0.5em}
    \includegraphics[width=0.88\linewidth]{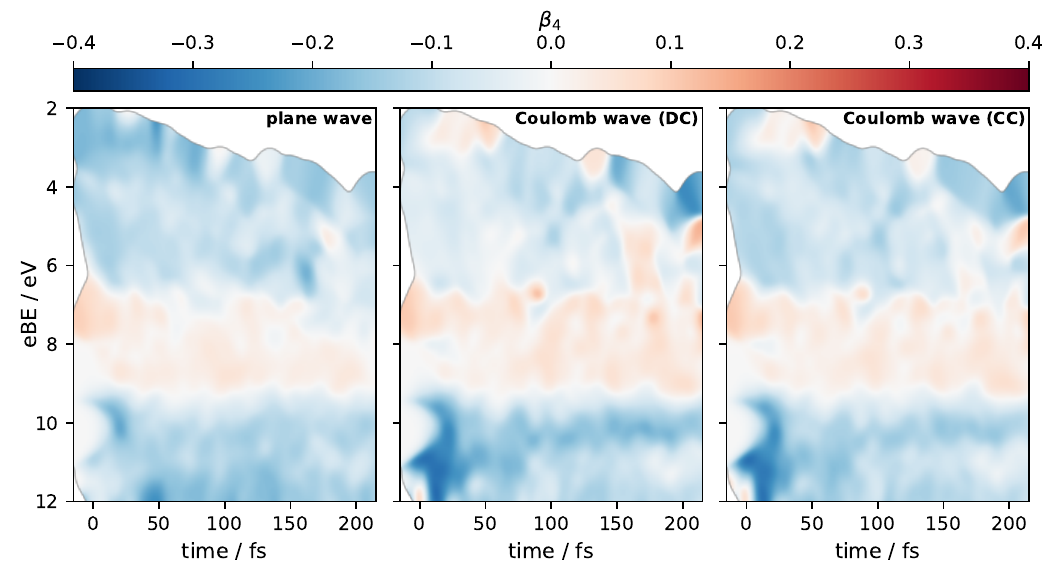}
    \caption{Time-resolved photoelectron anisotropy parameters $\beta_2$ (upper
        row) and $\beta_4$ (lower row) for the pre-aligned ensemble, obtained from
        plane-wave matrix elements and from Coulomb-wave matrix elements using either
        the Dyson centroid (DC) or the charge center (CC) as the Coulomb-wave origin.
        The underlying ionization events were computed along the FISH dynamics
        trajectories using SA5-RMS-CASPT2(10, 10) and SA8-RMS-CASPT2(9, 10). The
        discrete anisotropy parameters were broadened using the same two-dimensional
        Gaussian profile as in
        Fig.~\ref{fig:trpes}, with $\sigma_t = \SI{4}{\fs}$ and
        $\sigma_{\mathrm{eBE}} = \SI{0.2}{\eV}$, by taking the ratio of the broadened
        anisotropy-weighted intensity to the broadened intensity. Values are shown only
        where the broadened intensity exceeds $10^{-4}$ of its maximum.}
    \label{fig:trpad}
\end{figure*}

The simulated signals in regions J, K and L contain photoproduct contributions.
Experimentally, the positive late-time intensity in region K is not clearly
resolved as a component distinct from H and E. A separation of the simulated
TRPES according to the photoproducts reveals minor differences as shown in
Fig.~\ref{fig:si-product-separated-trpes} of the Supporting Information. Dewar
furan contributes to signal K (D\textsubscript{0/1}) and the high energy part
of signal L (D\textsubscript{3}). The ring-puckering motion causes the eBE of
$\text{S}_0 \to \text{D}_0$ to oscillate, thus contributing to signal J during
the formation of Dewar furan. CPCA contributes to signals K
(D\textsubscript{1}) and L (D\textsubscript{2}), while ionization band for
$\text{S}_0 \to \text{D}_0$ is hidden under the ground state bleach signal. The
BDA sub-ensemble contains only one trajectory and is therefore not used to
assign distinct spectral features. Due to the structural flexibility of the
open-chained en-diyl, its signal has the most complex features. During the
ring-opening process, it contributes to signal J (D\textsubscript{0}) and K
(D\textsubscript{1}). Other cationic states are energetically too high to be
captured by the eBE range shown here. The fully opened structure, on the other
hand, contributes to all three signals through at least ionization channels
involving D\textsubscript{0} through D\textsubscript{4}.

The time-dependent $\beta_2$ maps provide a more sensitive test of the
continuum approximation than the angle-integrated TRPES. The time-resolved
$\beta_2$ maps in Fig.~\ref{fig:trpad} were obtained using the same temporal
and energetic broadening as the TRPES. The corresponding isotropically averaged
and pre-aligned maps for all three continuum-wave treatments are compared in
Fig.~\ref{fig:si-trpad-averaging-comparison} of the Supporting Information.
Although the plane-wave and Coulomb-wave matrix elements give similar
angle-integrated TRPES, their $\beta_2$ maps differ substantially. The
plane-wave calculation yields predominantly negative $\beta_2$ values over the
plotted time--eBE range, whereas both Coulomb-wave calculations yield mostly
positive $\beta_2$ values. Differences between the Dyson-centered and
charge-centered Coulomb waves, which are barely visible in the angle-integrated
TRPES, are also resolved in the $\beta_2$ maps. This confirms for the
time-resolved spectra the conclusion drawn from the static PES: the
angle-integrated intensity is comparatively insensitive to details of the
continuum description, while the photoelectron angular distribution is much
more sensitive. The corresponding $\beta_4$ maps for the pre-aligned ensemble
are shown in the lower row of Fig.~\ref{fig:trpad}.

A qualitative comparison can be made with the anisotropy maps reported by
Humeniuk \emph{et al.}\cite{humeniuk_time-resolved_2013}. Although their probe
photon energy was different, they assigned the high-anisotropy component near
$\mathrm{PKE}\approx\SI{1.8}{\eV}$ to ionization from a Rydberg state to
D\textsubscript{0}. They further related the time dependence of the anisotropy
to the ultrafast decay from the initially populated S\textsubscript{2} state to
S\textsubscript{1} and to the subsequent change of the Dyson orbital between
Rydberg-like and valence character. The increase of $\beta_2$ near
$\mathrm{eBE}\approx\SI{3}{\eV}$ in the present Coulomb-wave maps is consistent
with this interpretation, since the corresponding signal A is assigned to
ionization into D\textsubscript{0} with substantial Rydberg-state contribution.

\begin{figure}[!htbp]
    \centering
    \includegraphics[width=\linewidth]{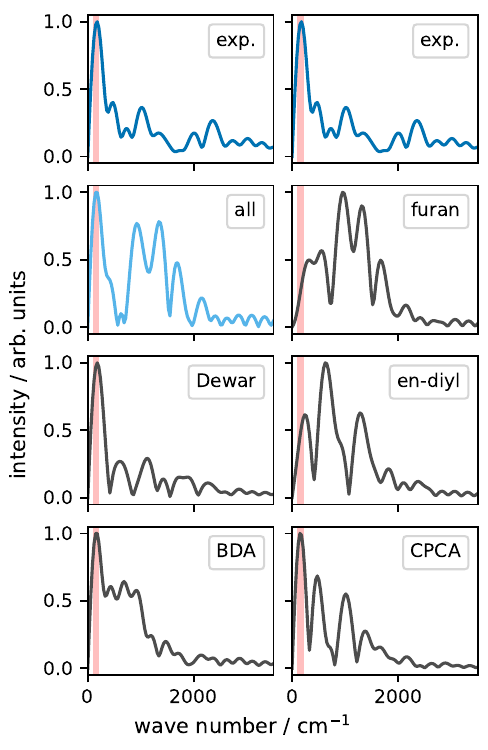}
    \caption{Fourier spectra of the integrated signal J from the experimental
        TRPES (exp.), the simulated Coulomb-wave TRPES (all), and simulated
        trajectory sub-ensembles separated by final product (furan, Dewar,
        en-diyl, BDA, and CPCA). Each spectrum is normalized to its maximum
        intensity. The shaded region spans \SIrange{100}{220}{\per\cm}, centered
        at \SI{160}{\per\cm}.}
    \label{fig:prod_freq}
\end{figure}

It is difficult to extract quantitative product ratios from the TRPES because
the photoproducts show strongly overlapping spectral features and the absolute
photoionization intensities depend on the continuum approximation. However, the
oscillatory behavior of signal J provides additional information about the
products formed. To analyze this contribution, the eBE range of signal J was
integrated for the experimental TRPES, the simulated Coulomb-wave TRPES, and
the product-separated simulated TRPES. Fourier transformation of the resulting
time series, using zero-padding, yields Fig.~\ref{fig:prod_freq}.

While the sub-ensemble reforming furan does not feature an oscillation at
around \SI{160}{\per\cm}, this component is present in both the experimental
and simulated TRPES. The formation of new photoproducts is therefore supported
by the frequency analysis. Because this frequency band is present in the TRPES
of sub-ensembles leading to Dewar furan, BDA, and CPCA, these products are
reasonable candidates upon excitation of furan.

\section{Conclusions}

In this work, we present a general theoretical framework for elucidating
ultrafast photochemical mechanisms from fs-EUV-TRPES. The framework combines
explicit-pulse nonadiabatic dynamics on dynamically correlated multireference
potential-energy surfaces with direct simulations of time-resolved
photoelectron observables. We apply this framework to the photoexcited
relaxation of gas-phase furan at \SI{200}{\nm}, using RMS-CASPT2 electronic
structure theory and field-induced surface hopping (FISH) dynamics.

First, electronic-structure benchmarks establish that
SA5-RMS-CASPT2(10,10)/aug-cc-pVDZ reproduces the excitation energies of the
first three singlet excited states in good agreement with high-quality
reference data. The active space contains all five \textpi-orbitals, the 3s
Rydberg orbital, and the four $\mathrm{\sigma}_{\text{CO}}^{(*)}$ orbitals,
providing a balanced description of the Franck--Condon region and the
ring-puckering and ring-opening minimum-energy conical intersections (MECIs).

Explicit treatment of the laser--molecule interaction shows that the
\SI{200}{\nm}, \SI{10}{\fs} pulse excites 55 of 250 Wigner-sampled trajectories
and prepares a distribution over the first three singlet excited states.
Sequential-exponential fits yield lifetimes of \SI{13}{\fs} for $\mathrm{S_3
        \to S_2}$, \SI{13}{\fs} for $\mathrm{S_2 \to S_1}$, and \SI{64}{\fs} for
$\mathrm{S_1 \to S_0}$ internal conversion. Approximately \SI{87}{\percent} of
the ground-state hops occur near the ring-puckering CI. Vibrational energy is
subsequently transferred from ring-puckering to ring-opening coordinates, and
ring-opening motion gains energy after about \SI{20}{\fs}. This helps reconcile
predominantly puckering-mediated internal conversion with signatures of
ring-opening motion in recent time-resolved studies.

At \SI{215}{\fs}, the trajectory ensemble contains predominantly hot furan,
alongside Dewar furan, the en-diyl biradical, cycloprop-2-ene-1-carbaldehyde
(CPCA), and buta-2,3-dienal (BDA). The predominance of hot furan and the
formation of Dewar furan and CPCA are consistent with \SI{200}{\nm}
time-resolved photoelectron measurements, while the open-chain products are
consistent with C--O bond-fission signatures from attosecond soft-X-ray
absorption and time-resolved Coulomb-explosion imaging. The low computed BDA
yield may explain its absence from ultrafast measurements, whereas secondary
rearrangement of en-diyl may impede its direct detection.

The simulated TRPES reproduces the main experimental features, including the
\SI{7.5}{\eV} band associated with ionization into D\textsubscript{2} and the
A--B and C--D kinks in the electron-binding-energy traces. The simulation also
predicts photoproduct contributions in regions J, K, and L. Evaluating
photoionization intensities with plane-wave or Coulomb-wave continuum functions
reduces the overly intense, slowly shifting contribution to signal A obtained
from squared Dyson norms. Although the continuum treatments give similar
angle-integrated spectra, their time-resolved $\beta_2$ maps differ
substantially, demonstrating that photoelectron anisotropy is more sensitive to
the continuum description than the angle-integrated TRPES. The oscillatory
component of signal J near \SI{160}{\per\cm} provides further support for the
formation of new photoproducts.

\begin{acknowledgement}
    The authors thank the Deutsche Forschungsgemeinschaft for the financial support
    \emph{via} Grant MI1236/7-1.
    XM thanks the Studienstiftung des deutschen Volkes e.V. for a fellowship.
\end{acknowledgement}

\begin{suppinfo}
    Optimized geometries, electronic-structure benchmarks, trajectory analyses,
    and supplementary photoelectron spectra and anisotropies.
\end{suppinfo}

\bibliography{bibliography,bibliography-utf}

\end{document}

\maketitle
\tableofcontents
\vfill
\clearpage

\section{Equilibrium geometries and vertical excitation energies}
\label{sec:si-equilibrium-geometries}

\begin{table}[H]
    \centering
    \caption*{SA5-RMS-CASPT2(10,10)/aug-cc-pVDZ equilibrium geometry}
    \begin{tabular}{c S[table-format=+1.6] S[table-format=+1.6] S[table-format=+1.6]}
        \hline
          & \multicolumn{1}{c}{$x$ / \si{\angstrom}} & \multicolumn{1}{c}{$y$ / \si{\angstrom}} & \multicolumn{1}{c}{$z$ / \si{\angstrom}} \\
        \hline
        O & 0.000000                                 & 0.000000                                 & 1.145536                                 \\
        C & 0.000000                                 & 1.102844                                 & 0.323348                                 \\
        C & 0.000000                                 & 0.723373                                 & -0.999206                                \\
        C & 0.000000                                 & -0.723373                                & -0.999206                                \\
        C & 0.000000                                 & -1.102844                                & 0.323348                                 \\
        H & 0.000000                                 & 2.067330                                 & 0.823018                                 \\
        H & 0.000000                                 & 1.388218                                 & -1.860704                                \\
        H & 0.000000                                 & -1.388218                                & -1.860704                                \\
        H & 0.000000                                 & -2.067330                                & 0.823018                                 \\
        \hline
    \end{tabular}
\end{table}

\begin{table}[H]
    \centering
    \caption*{CCSD(T)/cc-pVTZ equilibrium geometry}
    \begin{tabular}{c S[table-format=+1.6] S[table-format=+1.6] S[table-format=+1.6]}
        \hline
          & \multicolumn{1}{c}{$x$ / \si{\angstrom}} & \multicolumn{1}{c}{$y$ / \si{\angstrom}} & \multicolumn{1}{c}{$z$ / \si{\angstrom}} \\
        \hline
        O & 0.000000                                 & 0.000000                                 & 1.136645                                 \\
        C & 0.000000                                 & 1.093069                                 & 0.319018                                 \\
        C & 0.000000                                 & 0.719743                                 & -0.989791                                \\
        C & 0.000000                                 & -0.719743                                & -0.989791                                \\
        C & 0.000000                                 & -1.093069                                & 0.319018                                 \\
        H & 0.000000                                 & 2.048229                                 & 0.814154                                 \\
        H & 0.000000                                 & 1.378486                                 & -1.841873                                \\
        H & 0.000000                                 & -1.378486                                & -1.841873                                \\
        H & 0.000000                                 & -2.048229                                & 0.814154                                 \\
        \hline
    \end{tabular}
\end{table}

\begin{table}[H]
    \centering
    \caption{Vertical excitation energies calculated using SA5-RMS-CASPT2(10,10)/aug-cc-pVDZ at the SA5-RMS-CASPT2(10,10)/aug-cc-pVDZ and CCSD(T)/cc-pVTZ equilibrium geometries.}
    \label{tab:si-vertical-excitation-energies}
    \begin{tabular}{l S[table-format=1.2] S[table-format=1.2]}
        \hline
        state                            & \multicolumn{1}{c}{SA5-RMS-CASPT2}                & \multicolumn{1}{c}{CCSD(T)}                       \\
                                         & \multicolumn{1}{c}{geometry / \si{\electronvolt}} & \multicolumn{1}{c}{geometry / \si{\electronvolt}} \\
        \hline
        $1\,^1A_2 (\pi \to \mathrm{3s})$ & 5.83                                              & 5.86                                              \\
        $1\,^1B_2 (\pi \to \pi^*)$       & 5.98                                              & 6.11                                              \\
        $2\,^1A_1 (\pi \to \pi^*)$       & 6.04                                              & 6.19                                              \\
        \hline
    \end{tabular}
\end{table}

\section{Exclusion of 3p Rydberg states}
\label{sec:si-3p-rydberg-exclusion}

The $2\,^1A_1$ state contains an appreciable admixture of doubly excited
configurations. Its position relative to the 3p Rydberg manifold is therefore
sensitive to how well the electronic-structure method treats correlation
involving the doubly excited component. As summarized in
Table~\ref{tab:si-3p-rydberg-energies}, CASPT2, NEVPT2, MRMP, and CC3 all place
$2\,^1A_1$ below the lowest 3p state.

Although no experimental excitation energy is available for $2\,^1A_1$, the
experimental 3p assignments provide an independent benchmark for the upper part
of this state manifold: the first 3p transition is observed at
\SI{6.47}{\electronvolt}. Its agreement with the calculated 3p energies lends
confidence to the theoretical ordering in which $2\,^1A_1$ precedes the 3p
manifold. The 3p states therefore lie above the three states targeted here, and
their explicit inclusion is not required for the electronic-state manifold
considered in this work.

\begin{table}[H]
    \centering
    \small
    \caption{Theoretical and experimental vertical excitation energies of the first six singlet excited states of furan. Energies are in \si{\electronvolt}. The present SA5-RMS-CASPT2 calculations do not include the 3p Rydberg states.}
    \label{tab:si-3p-rydberg-energies}
    \begin{tabular}{l c c c c c c}
        \hline
        state                              & this work & CASPT2\textsuperscript{a} & NEVPT2\textsuperscript{b} & MRMP\textsuperscript{c} & CC3\textsuperscript{d} & Exp.\textsuperscript{e} \\
        \hline
        $1\,^1A_2 (\pi \to \mathrm{3s})$   & 5.86      & 5.92                      & 6.13                      & 5.84                    & 6.08                   & 5.91                    \\
        $1\,^1B_2 (\pi \to \pi^*)$         & 6.11      & 6.04                      & 6.41                      & 5.95                    & 6.34                   & 6.04                    \\
        $2\,^1A_1 (\pi \to \pi^*)$         & 6.19      & 6.16                      & 6.62                      & 6.16                    & 6.58                   & --                      \\
        $1\,^1B_1 (\pi \to \mathrm{3p}_y)$ & --        & 6.48                      & 6.68                      & 6.40                    & 6.63                   & 6.47                    \\
        $2\,^1A_2 (\pi \to \mathrm{3p}_x)$ & --        & 6.61                      & 6.79                      & 6.53                    & 6.80                   & 6.61                    \\
        $2\,^1B_2 (\pi \to \mathrm{3p}_z)$ & --        & 6.48                      & 6.87                      & 6.51                    & 7.23                   & 6.75                    \\
        \hline
    \end{tabular}

    \vspace{0.5em}
    \footnotesize
    \textsuperscript{a} From Ref.~\citenum{serrano-andres_theoretical_1993}.
    \textsuperscript{b} PC-QDNEVPT2 results for the (0805) and (4322) active
    spaces from Ref.~\citenum{pastore_application_2006}.
    \textsuperscript{c} From Ref.~\citenum{christiansen_electronic_1998}.
    \textsuperscript{d} From Ref.~\citenum{loos_mountaineering_2020}.
    \textsuperscript{e} Summarized by Wan \emph{et al.}\cite{wan_electronic_2000}
    from Refs.~\citenum{pickett_vibrational_1940,price_absorption_1941,pickett_vacuum_1951,derrick_photoelectron-spectroscopical_1971,robin_higher_1975,flicker_electron_1976,flicker_trlplet_1976,cooper_resonantly_1980,roebber_electronic_1980,palmer_electronic_1995}.
\end{table}

\section{Minimum-energy conical-intersection geometries}
\label{sec:si-meci-geometries}

\begin{table}[H]
    \centering
    \caption*{Ring-opening minimum-energy conical intersection (MECI\textsubscript{O})}
    \begin{tabular}{c S[table-format=+1.6] S[table-format=+1.6] S[table-format=+1.6]}
        \hline
          & \multicolumn{1}{c}{$x$ / \si{\angstrom}} & \multicolumn{1}{c}{$y$ / \si{\angstrom}} & \multicolumn{1}{c}{$z$ / \si{\angstrom}} \\
        \hline
        O & 0.000410                                 & -0.602481                                & 1.759937                                 \\
        C & -0.000010                                & 1.536915                                 & 0.460370                                 \\
        C & -0.000204                                & 0.792656                                 & -0.658519                                \\
        C & 0.000042                                 & -0.662384                                & -0.596411                                \\
        C & -0.000180                                & -1.276096                                & 0.697093                                 \\
        H & -0.000041                                & 2.592734                                 & 0.732202                                 \\
        H & -0.000033                                & 1.273623                                 & -1.649455                                \\
        H & 0.000228                                 & -1.274061                                & -1.501016                                \\
        H & -0.000212                                & -2.380907                                & 0.755800                                 \\
        \hline
    \end{tabular}
\end{table}

\begin{table}[H]
    \centering
    \caption*{Ring-puckering minimum-energy conical intersection (MECI\textsubscript{P})}
    \begin{tabular}{c S[table-format=+1.6] S[table-format=+1.6] S[table-format=+1.6]}
        \hline
          & \multicolumn{1}{c}{$x$ / \si{\angstrom}} & \multicolumn{1}{c}{$y$ / \si{\angstrom}} & \multicolumn{1}{c}{$z$ / \si{\angstrom}} \\
        \hline
        O & -0.840736                                & 0.095195                                 & 1.336802                                 \\
        C & -0.067293                                & 0.894378                                 & 0.676100                                 \\
        C & 0.041345                                 & 0.606483                                 & -0.788997                                \\
        C & -0.107732                                & -0.744522                                & -0.819726                                \\
        C & -0.251949                                & -1.268806                                & 0.605693                                 \\
        H & 0.449588                                 & 1.700707                                 & 1.216608                                 \\
        H & 0.130585                                 & 1.351168                                 & -1.583373                                \\
        H & -0.127106                                & -1.398370                                & -1.694040                                \\
        H & 0.773298                                 & -1.236232                                & 1.050934                                 \\
        \hline
    \end{tabular}
\end{table}

\clearpage
\section{Natural orbitals of the enlarged cationic active space}
\label{sec:si-sa10-natural-orbitals}

\begin{figure}[H]
    \centering
    \includegraphics[width=\linewidth]{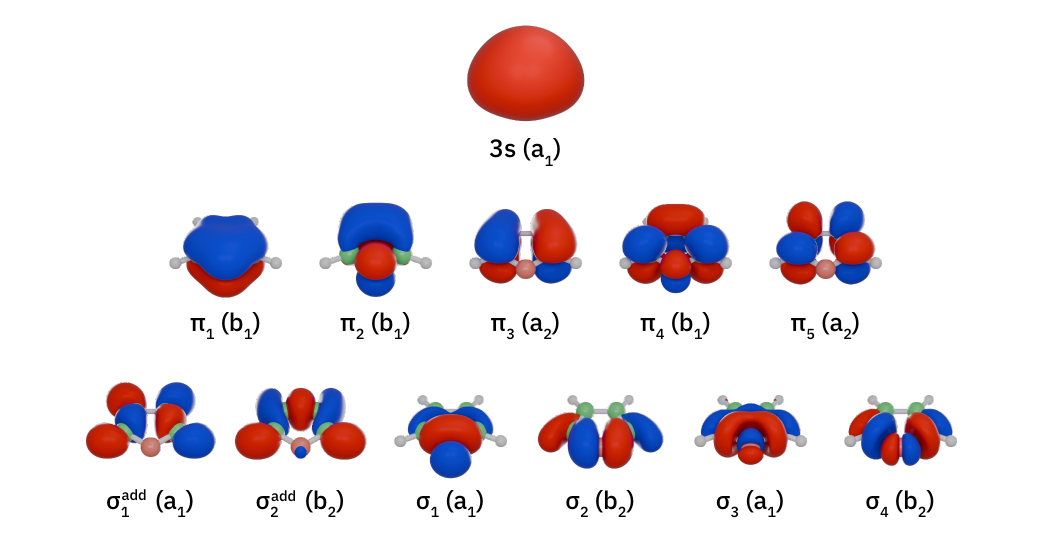}
    \caption{Natural orbitals of furan at the SA10-RMS-CASPT2(13,12)/aug-cc-pVDZ level. An isovalue of 0.04 is used for the valence orbitals and 0.01 for the Rydberg orbital. The 3s Rydberg orbital is displayed at a reduced scale for clarity.}
    \label{fig:si-sa10-natural-orbitals}
\end{figure}

\section{Single-point and ensemble photoelectron spectra}
\label{sec:si-ensemble-pes}

\begin{figure}[H]
    \centering
    \includegraphics[width=0.9\linewidth]{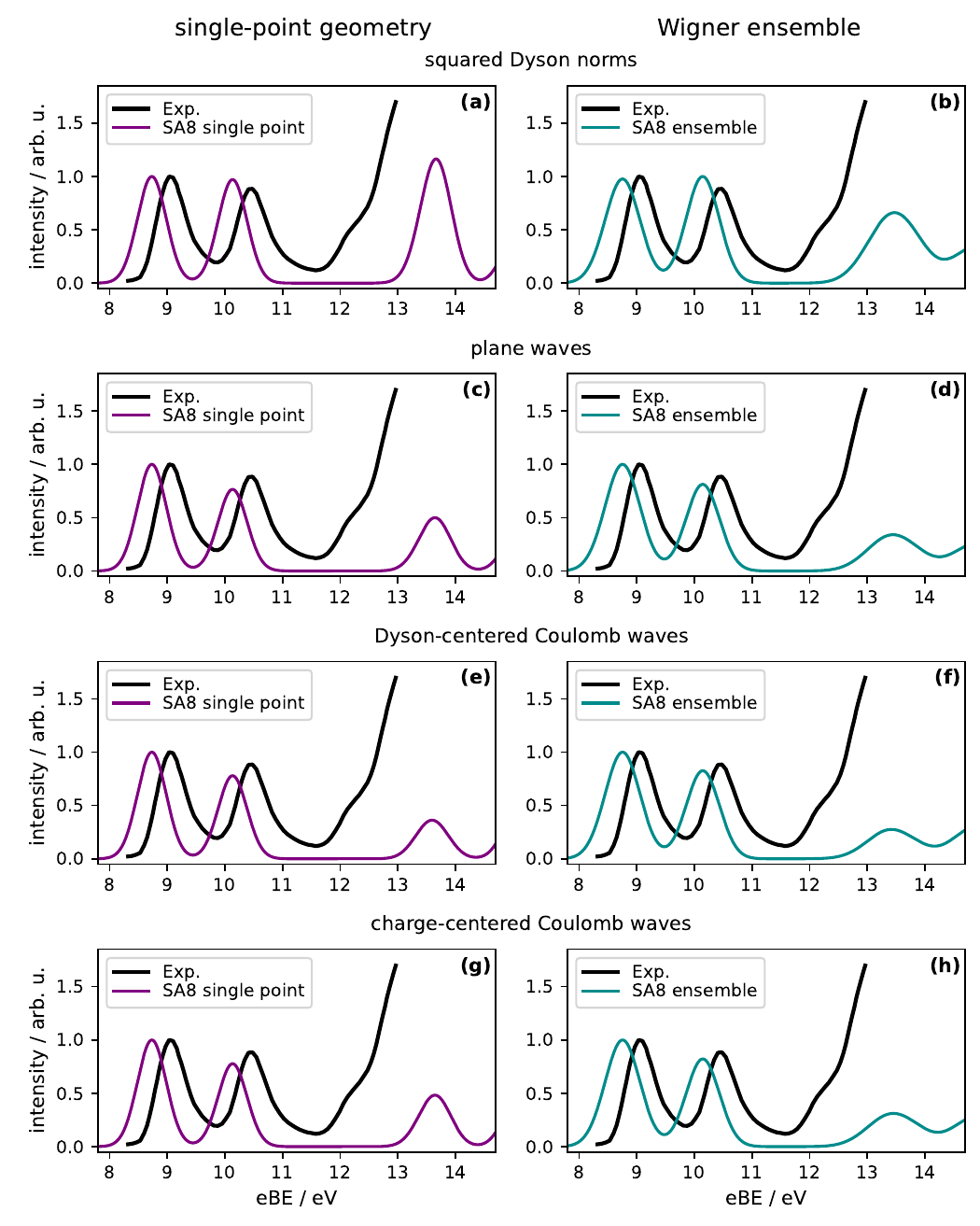}
    \caption{Comparison of the experimental photoelectron spectrum (PES) of
        furan reproduced from Ref.~\citenum{uenishi_signatures_2024} with
        SA8-RMS-CASPT2(9, 10) spectra calculated at a single reference geometry
        (left column) and for an ensemble of 250 geometries sampled from a
        harmonic canonical Wigner distribution at \SI{50}{\K} (right column).
        The relative photoionization intensities were estimated using
        (a, b) squared Dyson norms, (c, d) plane-wave matrix elements,
        (e, f) Dyson-centered Coulomb-wave matrix elements, and
        (g, h) charge-centered Coulomb-wave matrix elements.
        A Gaussian broadening with $\sigma=\SI{0.25}{\eV}$ was applied
        to all calculated spectra.}
    \label{fig:si-ensemble-pes}
\end{figure}

\section{Single-point and ensemble photoelectron anisotropies}
\label{sec:si-ensemble-beta2}

\begin{figure}[H]
    \centering
    \includegraphics[width=0.9\linewidth]{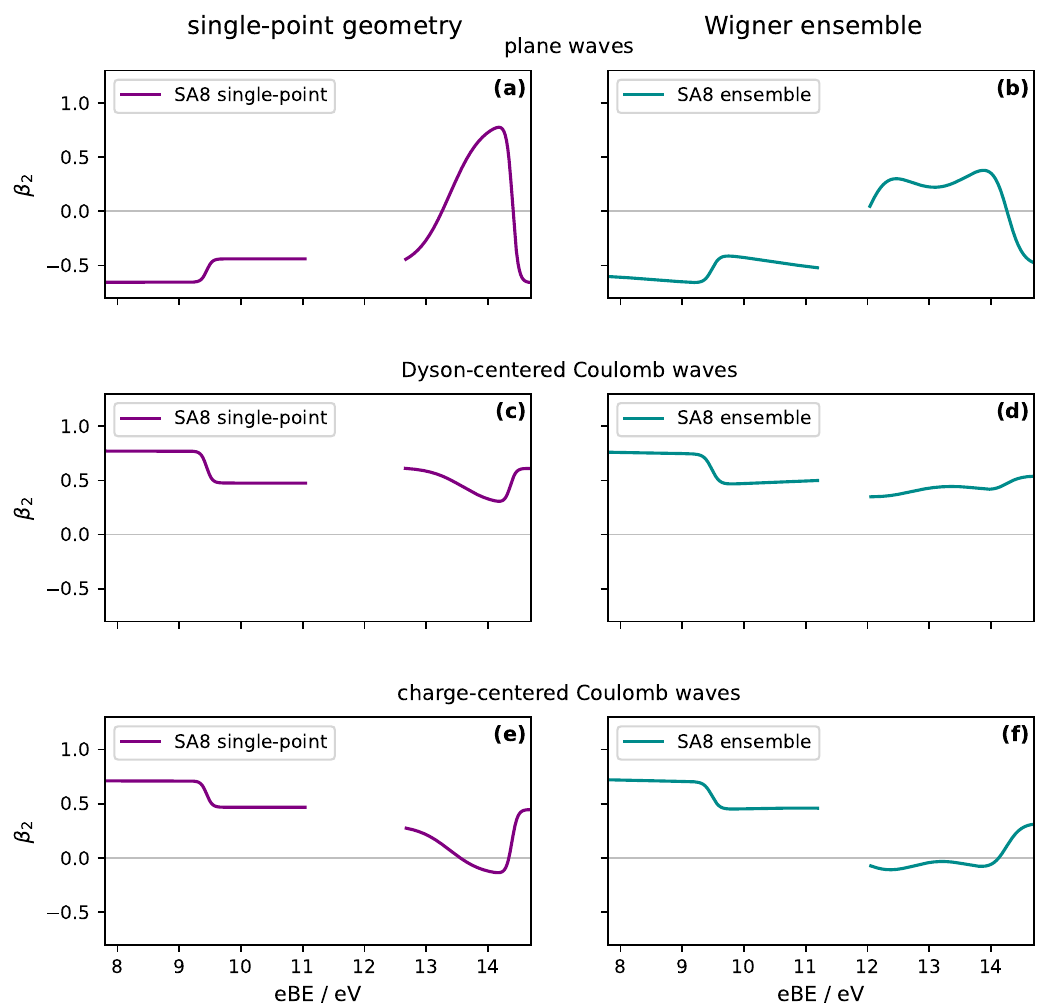}
    \caption{Comparison of the simulated photoelectron anisotropy parameter $\beta_2$ of
        furan employing SA8-RMS-CASPT2(9, 10) calculated at a single reference geometry
        (left column) and for an ensemble of 250 geometries sampled from a
        harmonic canonical Wigner distribution at \SI{50}{\K} (right column).
        The anisotropy parameters were estimated using
        (a, b) plane-wave matrix elements,
        (c, d) Dyson-centered Coulomb-wave matrix elements, and
        (e, f) charge-centered Coulomb-wave matrix elements.
        The discrete $\beta_2$ values were broadened with Gaussians of
        $\sigma=\SI{0.25}{\eV}$ by taking the ratio of the broadened
        anisotropy-weighted intensity to the broadened intensity.
        Values are shown only where the broadened intensity exceeds $10^{-3}$
        of its maximum.}
    \label{fig:si-ensemble-beta2}
\end{figure}
\clearpage

\section{Classification of nonadiabatic relaxation pathways}
\label{sec:si-trajectory-classification}

Each hop geometry was aligned with the two reference MECI structures while
accounting for symmetry-equivalent orientations. The mass-weighted RMSD to each
reference was then evaluated, and the geometry was assigned to
MECI\textsubscript{P} or MECI\textsubscript{O} according to the smaller RMSD.
The same procedure was applied to the geometry at which the
S\textsubscript{1}--S\textsubscript{0} energy gap reached its minimum along
each trajectory.

\begin{table}[H]
    \centering
    \caption{Classification of the 53 trajectories that returned to
        S\textsubscript{0}. Fractions are given with 95\% confidence intervals,
        and times are reported as mean $\pm$ standard deviation.}
    \label{tab:si-trajectory-classification}
    \renewcommand{\arraystretch}{1.3}
    \begin{tabular}{lccc S[table-format=3.0(2)]}
        \hline
        geometry used for classification & assignment            & count & fraction / \%    & \multicolumn{1}{c}{time / \si{\fs}} \\
        \hline
        hop geometry                     & MECI\textsubscript{P} & 46    & $87^{+8}_{-12}$  & 82 \pm 47                           \\
        hop geometry                     & MECI\textsubscript{O} & 7     & $13^{+12}_{-8}$  & 84 \pm 45                           \\
        minimum-gap geometry             & MECI\textsubscript{P} & 30    & $57^{+13}_{-15}$ & 109 \pm 47                          \\
        minimum-gap geometry             & MECI\textsubscript{O} & 23    & $43^{+15}_{-13}$ & 111 \pm 54                          \\
        \hline
    \end{tabular}
\end{table}

\section{Trajectory-resolved relaxation pathways and final structures}
\label{sec:si-trajectory-details}

No hop or minimum-gap data are available for trajectories 0229 and 0232, which
did not return to S\textsubscript{0} at the end of the simulation. For these
two trajectories, "Furan" denotes a furan-like final nuclear geometry rather
than a ground-state photoproduct.

\begingroup
\small
\setlength{\tabcolsep}{5pt}
\begin{longtable}{rccrrl}
    \caption{Trajectory-resolved MECI assignments, characteristic times, and
        final structural classifications for all 55 excited trajectories.}
    \label{tab:si-trajectory-details}                                                                                                       \\
    \hline
    traj. & hop assignment        & min.-gap assignment   & $t_{\mathrm{hop}}$ / \si{\fs} & $t_{\mathrm{min}}$ / \si{\fs} & final structure \\
    \hline
    \endfirsthead
    \multicolumn{6}{c}{Table~\thetable\ continued}                                                                                          \\
    \hline
    traj. & hop assignment        & min.-gap assignment   & $t_{\mathrm{hop}}$ / \si{\fs} & $t_{\mathrm{min}}$ / \si{\fs} & final structure \\
    \hline
    \endhead
    \hline
    \multicolumn{6}{r}{Continued on the next page}                                                                                          \\
    \endfoot
    \hline
    \endlastfoot
    0000  & MECI\textsubscript{P} & MECI\textsubscript{P} & 95.3                          & 98.1                          & Dewar           \\
    0001  & MECI\textsubscript{P} & MECI\textsubscript{O} & 33.3                          & 107.7                         & Furan           \\
    0003  & MECI\textsubscript{P} & MECI\textsubscript{P} & 73.9                          & 80.5                          & CPCA            \\
    0015  & MECI\textsubscript{P} & MECI\textsubscript{P} & 35.5                          & 35.9                          & en-diyl         \\
    0016  & MECI\textsubscript{P} & MECI\textsubscript{P} & 66.7                          & 56.7                          & Dewar           \\
    0018  & MECI\textsubscript{O} & MECI\textsubscript{O} & 67.5                          & 62.1                          & Furan           \\
    0019  & MECI\textsubscript{P} & MECI\textsubscript{O} & 127.3                         & 198.3                         & Furan           \\
    0024  & MECI\textsubscript{P} & MECI\textsubscript{P} & 170.9                         & 175.5                         & Furan           \\
    0026  & MECI\textsubscript{O} & MECI\textsubscript{O} & 35.3                          & 35.7                          & en-diyl         \\
    0027  & MECI\textsubscript{P} & MECI\textsubscript{O} & 42.3                          & 167.1                         & Furan           \\
    0028  & MECI\textsubscript{P} & MECI\textsubscript{P} & 191.3                         & 190.5                         & Dewar           \\
    0035  & MECI\textsubscript{P} & MECI\textsubscript{P} & 176.1                         & 165.5                         & Furan           \\
    0037  & MECI\textsubscript{P} & MECI\textsubscript{P} & 93.5                          & 96.5                          & Furan           \\
    0044  & MECI\textsubscript{P} & MECI\textsubscript{P} & 56.1                          & 58.9                          & Furan           \\
    0049  & MECI\textsubscript{P} & MECI\textsubscript{P} & 69.7                          & 69.9                          & Furan           \\
    0055  & MECI\textsubscript{P} & MECI\textsubscript{O} & 43.1                          & 70.5                          & Furan           \\
    0058  & MECI\textsubscript{P} & MECI\textsubscript{P} & 87.3                          & 87.7                          & Furan           \\
    0059  & MECI\textsubscript{P} & MECI\textsubscript{P} & 78.3                          & 83.7                          & Furan           \\
    0066  & MECI\textsubscript{P} & MECI\textsubscript{P} & 62.5                          & 62.9                          & Furan           \\
    0075  & MECI\textsubscript{O} & MECI\textsubscript{O} & 167.3                         & 130.1                         & CPCA            \\
    0076  & MECI\textsubscript{P} & MECI\textsubscript{P} & 148.1                         & 91.3                          & Furan           \\
    0077  & MECI\textsubscript{O} & MECI\textsubscript{O} & 52.3                          & 52.5                          & CPCA            \\
    0080  & MECI\textsubscript{P} & MECI\textsubscript{P} & 114.5                         & 114.7                         & Furan           \\
    0083  & MECI\textsubscript{P} & MECI\textsubscript{O} & 20.1                          & 56.9                          & Furan           \\
    0085  & MECI\textsubscript{P} & MECI\textsubscript{P} & 33.9                          & 42.1                          & Furan           \\
    0087  & MECI\textsubscript{P} & MECI\textsubscript{P} & 168.9                         & 169.3                         & Furan           \\
    0088  & MECI\textsubscript{P} & MECI\textsubscript{O} & 35.1                          & 60.1                          & Furan           \\
    0096  & MECI\textsubscript{P} & MECI\textsubscript{P} & 64.5                          & 64.9                          & Furan           \\
    0113  & MECI\textsubscript{P} & MECI\textsubscript{P} & 178.3                         & 178.7                         & en-diyl         \\
    0115  & MECI\textsubscript{P} & MECI\textsubscript{P} & 116.1                         & 114.5                         & Furan           \\
    0126  & MECI\textsubscript{P} & MECI\textsubscript{O} & 23.5                          & 185.3                         & Furan           \\
    0131  & MECI\textsubscript{P} & MECI\textsubscript{O} & 102.7                         & 144.3                         & Furan           \\
    0133  & MECI\textsubscript{P} & MECI\textsubscript{P} & 139.5                         & 140.1                         & Dewar           \\
    0138  & MECI\textsubscript{O} & MECI\textsubscript{O} & 62.9                          & 63.7                          & BDA             \\
    0139  & MECI\textsubscript{P} & MECI\textsubscript{O} & 29.1                          & 55.5                          & Furan           \\
    0140  & MECI\textsubscript{P} & MECI\textsubscript{P} & 48.9                          & 213.5                         & Furan           \\
    0145  & MECI\textsubscript{P} & MECI\textsubscript{O} & 19.1                          & 106.9                         & en-diyl         \\
    0152  & MECI\textsubscript{O} & MECI\textsubscript{O} & 137.1                         & 108.5                         & en-diyl         \\
    0171  & MECI\textsubscript{P} & MECI\textsubscript{O} & 15.7                          & 179.1                         & en-diyl         \\
    0172  & MECI\textsubscript{P} & MECI\textsubscript{O} & 116.9                         & 190.9                         & Furan           \\
    0173  & MECI\textsubscript{P} & MECI\textsubscript{P} & 78.9                          & 163.5                         & Furan           \\
    0177  & MECI\textsubscript{P} & MECI\textsubscript{P} & 91.1                          & 141.3                         & CPCA            \\
    0178  & MECI\textsubscript{P} & MECI\textsubscript{O} & 23.1                          & 53.5                          & Furan           \\
    0182  & MECI\textsubscript{P} & MECI\textsubscript{P} & 110.9                         & 103.1                         & Furan           \\
    0190  & MECI\textsubscript{P} & MECI\textsubscript{P} & 73.7                          & 74.7                          & Dewar           \\
    0196  & MECI\textsubscript{P} & MECI\textsubscript{P} & 107.9                         & 112.5                         & Dewar           \\
    0203  & MECI\textsubscript{P} & MECI\textsubscript{O} & 71.9                          & 170.3                         & Furan           \\
    0209  & MECI\textsubscript{P} & MECI\textsubscript{P} & 56.1                          & 60.5                          & Furan           \\
    0214  & MECI\textsubscript{O} & MECI\textsubscript{O} & 62.9                          & 63.3                          & CPCA            \\
    0215  & MECI\textsubscript{P} & MECI\textsubscript{P} & 90.1                          & 88.7                          & Dewar           \\
    0226  & MECI\textsubscript{P} & MECI\textsubscript{O} & 26.1                          & 101.1                         & Furan           \\
    0229  & --                    & --                    & --                            & --                            & Furan           \\
    0231  & MECI\textsubscript{P} & MECI\textsubscript{P} & 97.5                          & 134.1                         & Furan           \\
    0232  & --                    & --                    & --                            & --                            & Furan           \\
    0248  & MECI\textsubscript{P} & MECI\textsubscript{O} & 68.7                          & 177.3                         & Furan           \\
\end{longtable}
\endgroup

\clearpage
\section{Dimensionality reduction of the trajectory geometries}
\label{sec:si-trajectory-embedding}

The structural evolution of the 55 trajectories was visualized by embedding 116
frames per trajectory, sampled every \SI{2}{\fs}, together with the optimized
furan, MECI, and product geometries in two dimensions.

The first step towards applying the dimensionality-reduction techniques
discussed here is to define a measure of structural dissimilarity between
molecular geometries. Consider geometries $\tilde{\bm{X}}^{(i)}$ and
$\tilde{\bm{X}}^{(j)}$ from a set of $n$ geometries of the same molecule with
$N$ atoms, where $\tilde{\vec{x}}_a^{(i)}$ and $\tilde{\vec{x}}_a^{(j)}$ are
the raw Cartesian coordinate vectors of atom $a$. Before comparison, each
geometry is translated so that its center of mass lies at the origin:
\begin{equation}
    \vec{x}_a^{(i)}
    =\tilde{\vec{x}}_a^{(i)}
    -\frac{1}{M}\sum_{b=1}^{N}m_b\tilde{\vec{x}}_b^{(i)}\,,
    \qquad
    \vec{x}_a^{(j)}
    =\tilde{\vec{x}}_a^{(j)}
    -\frac{1}{M}\sum_{b=1}^{N}m_b\tilde{\vec{x}}_b^{(j)}\,,
\end{equation}
where $m_a$ is the mass of atom $a$ and $M=\sum_a m_a$. The centered
coordinate matrices $\bm{X}^{(i)}$ and $\bm{X}^{(j)}$ comprise the corresponding
centered coordinate vectors. Their mass-weighted, permutation- and
inversion-invariant root-mean-square deviation (RMSD) is defined as
\begin{equation}
    d_{ij} :=
    d\left(\bm{X}^{(i)},\bm{X}^{(j)}\right)
    = \min_{\bm{R},P,s}
    \left[
        \frac{1}{M}
        \sum_{a=1}^{N} m_a
        \left\lVert
        \vec{x}_a^{(i)}-s\bm{R}\vec{x}_{P(a)}^{(j)}
        \right\rVert^2
        \right]^{1/2}\,,
    \label{eq:si-structural-distance}
\end{equation}
where the proper rotation $\bm{R}\in\mathrm{SO}(3)$ was optimized using the
Kabsch algorithm\cite{kabsch1976solution}. The permutation $P$ spans the full
$S_4(\mathrm{C})\times S_4(\mathrm{H})$ group with oxygen fixed, and $s=\pm1$
includes inversion. The resulting distance is defined on molecular shape space
and need not be Euclidean. The maps below therefore represent structural
dissimilarity rather than physical coordinates.

Classical multidimensional scaling (CMDS), also known as principal coordinates
analysis (PCoA)\cite{borg2005modern}, constructs the centered Gram matrix
\begin{equation}
    \bm{B}=-\frac{1}{2}\bm{J}\bm{D}^{(2)}\bm{J}\quad\text{with}\quad
    \bm{J}=\mathbbm{1}-\frac{1}{n}\vec{1}\vec{1}^{\intercal} \text{ and } \left[\bm{D}^{(2)}\right]_{ij}=d_{ij}^2\,,
\end{equation}
where $\mathbbm{1}$ is the
$n\times n$ identity matrix, and $\vec{1}\vec{1}^{\intercal}$ is the matrix of
ones. The coordinates follow from the largest positive eigenvalues and
corresponding eigenvectors of $\bm{B}$. Negative eigenvalues indicate a
non-Euclidean dissimilarity matrix.

The representation used in Fig.~10(b) of the main manuscript was instead
obtained by metric multidimensional scaling (MDS)\cite{borg2005modern}. For
target distances $d_{ij}$ and two-dimensional coordinate vectors $\vec{z}_i$,
MDS minimizes the raw stress
\begin{equation}
    \mathcal{S}(\bm{Z})
    = \sum_{i<j}
    \left(d_{ij}-\lVert\vec{z}_i-\vec{z}_j\rVert\right)^2,
    \label{eq:si-mds-stress}
\end{equation}
using stress majorization (SMACOF), initialized with the CMDS coordinates. The
MDS axes have no individual physical meaning. Only relative positions are
interpreted.

To introduce temporal information, temporally weighted MDS minimizes the
weighted analogue of Eq.~\eqref{eq:si-mds-stress}. All pairs have unit weight,
while frames from the same trajectory and separated by a lag
$\tau\leq\tau_{\max}$ receive
\begin{equation}
    w(\tau)
    =1+A\exp\left[-\frac{\tau^2-1}{2\sigma^2}\right].
\end{equation}
We used $A=99$, $\sigma=3$ frames, and $\tau_{\max}=15$ frames. Thus,
distances between temporally nearby configurations are preferentially preserved.

The variational approach for Markov processes (VAMP) instead seeks coordinates
that retain dynamically predictive information\cite{wu_variational_2020}. Let
$\vec{\chi}_t$ and $\vec{\chi}_{t+\tau}$ denote the standardized 21-dimensional
metric-MDS feature vectors at the beginning and end of a lagged pair, with
respective means $\vec{\mu}_0$ and $\vec{\mu}_1$, and define
$\delta\vec{\chi}_0=\vec{\chi}_t-\vec{\mu}_0$ and
$\delta\vec{\chi}_1=\vec{\chi}_{t+\tau}-\vec{\mu}_1$. The instantaneous
covariance matrices and time-lagged cross-covariance matrix are
\begin{equation}
    \bm{C}_{00}=\left\langle\delta\vec{\chi}_0\delta\vec{\chi}_0^{\mathrm T}\right\rangle,
    \quad
    \bm{C}_{01}=\left\langle\delta\vec{\chi}_0\delta\vec{\chi}_1^{\mathrm T}\right\rangle,
    \quad
    \bm{C}_{11}=\left\langle\delta\vec{\chi}_1\delta\vec{\chi}_1^{\mathrm T}\right\rangle.
\end{equation}
The covariance-normalized Koopman matrix and its singular value decomposition
are given by
\begin{equation}
    \bar{\bm{K}}
    =\bm{C}_{00}^{-1/2}\bm{C}_{01}\bm{C}_{11}^{-1/2}
    =\bm{U}\bm{\Sigma}\bm{V}^{\mathrm T}.
\end{equation}
The first two left singular components were retained and scaled by their
singular values to obtain the kinetic-map coordinates\cite{noe_kinetic_2015}.
We used a lag of one frame (\SI{2}{\fs}), excluding pairs across trajectory
boundaries. The optimized reference geometries were projected only after fitting.

\begin{figure}[H]
    \centering
    \includegraphics[width=0.92\linewidth]{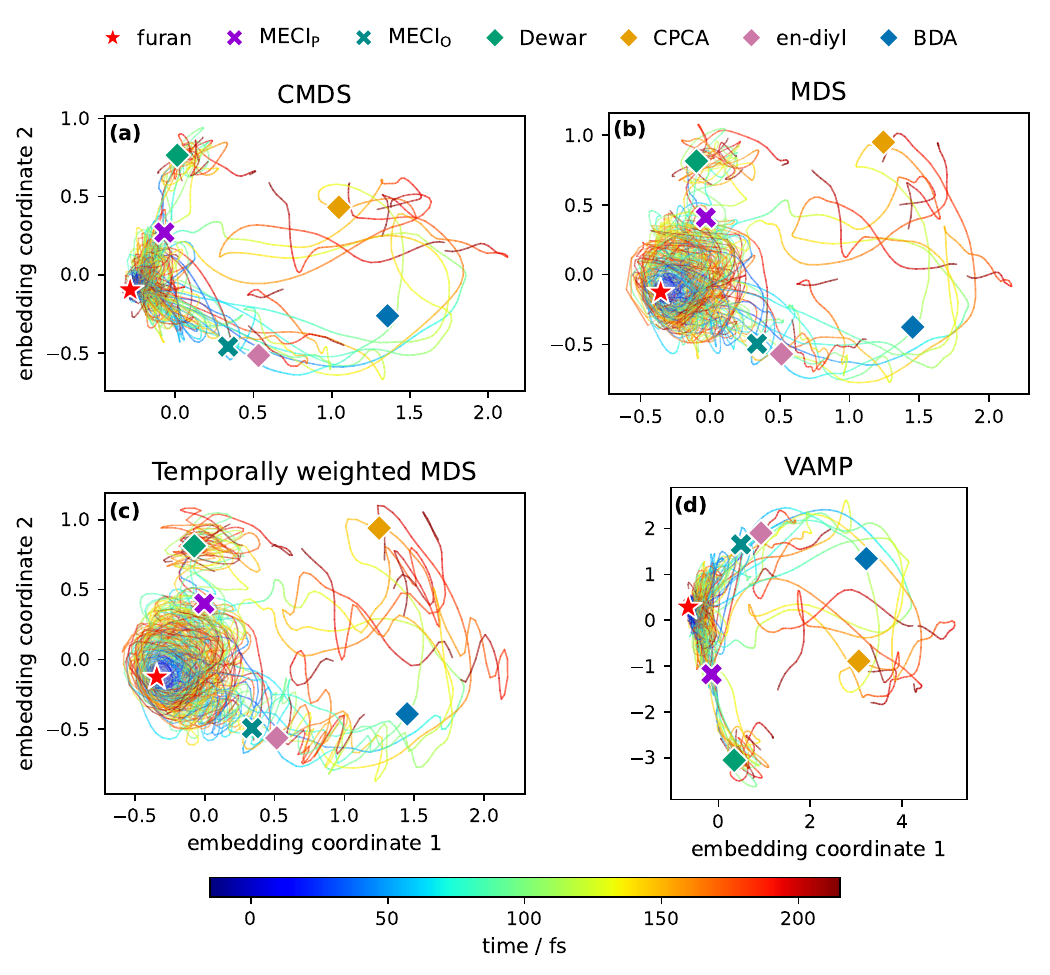}
    \caption{Comparison of two-dimensional embeddings of the trajectory and
        optimized reference geometries obtained using (a) classical
        multidimensional scaling (CMDS), (b) metric multidimensional scaling
        (MDS), (c) temporally weighted MDS, and (d) the variational approach for
        Markov processes (VAMP). Lines connect consecutive frames of individual
        trajectories and are colored according to simulation time. Symbols mark
        the optimized furan, MECI, and product reference geometries. The same
        inverse-density compensation is applied to the trajectory segments in
        all panels to enhance their visibility in sparsely populated regions.}
    \label{fig:si-embedding-comparison}
\end{figure}

Figure~\ref{fig:si-embedding-comparison} compares the four embeddings. CMDS and
MDS emphasize global structural relationships, temporally weighted MDS gives
additional weight to short-time evolution, and VAMP emphasizes dynamically
predictive variation. Nevertheless, all four maps show similar excursions from
the furan region towards the product reference structures.

\section{Effect of rotational averaging on the simulated TRPES}
\label{sec:si-trpes-averaging-comparison}

\begin{figure}[H]
    \centering
    \includegraphics[width=0.95\linewidth]{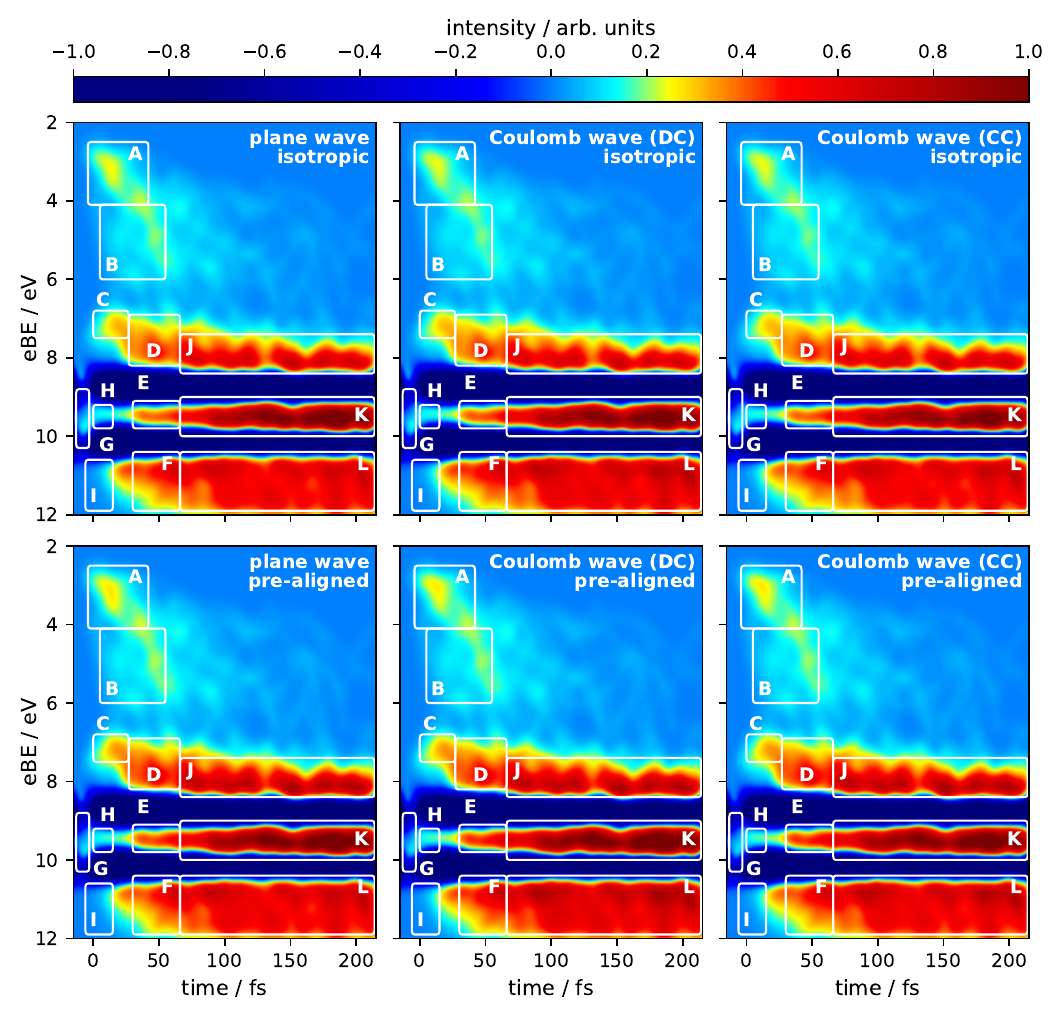}
    \caption{Comparison of simulated time-resolved photoelectron spectra
        obtained using isotropic rotational averaging (upper row) and
        pre-aligned rotational averaging (lower row). The relative
        photoionization intensities were calculated using plane-wave matrix
        elements (left column), Dyson-centered Coulomb-wave matrix elements
        (middle column), and charge-centered Coulomb-wave matrix elements
        (right column).}
    \label{fig:si-trpes-averaging-comparison}
\end{figure}

\section{TRPES separated by the initially populated electronic state}
\label{sec:si-excitation-separated-trpes}

\begin{figure}[H]
    \centering
    \includegraphics[width=0.95\linewidth]{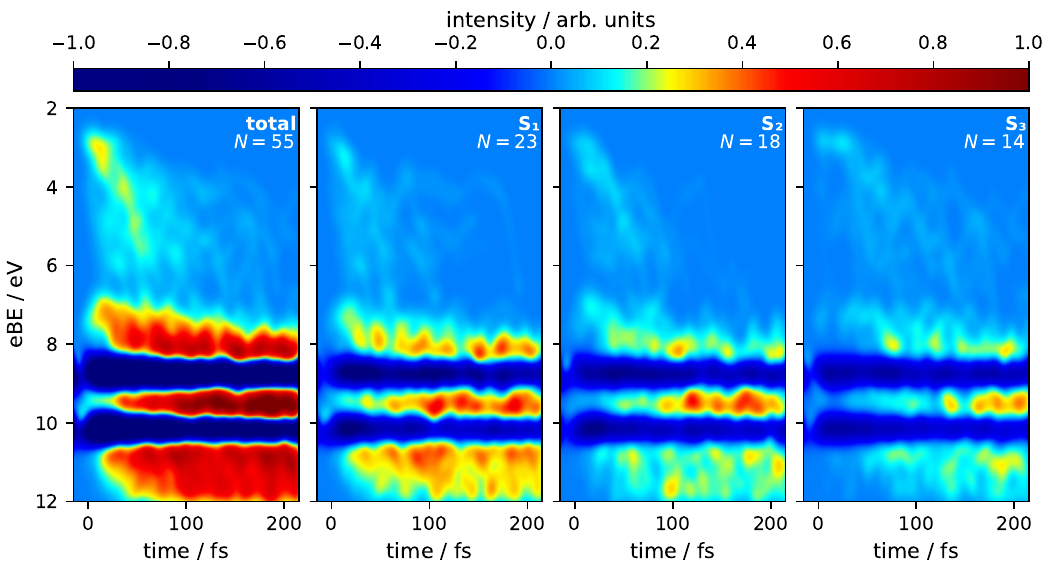}
    \caption{Simulated pre-aligned, Dyson-centered Coulomb-wave TRPES for the
        complete trajectory ensemble and separated according to the neutral
        electronic state occupied at zero pump--probe delay. The number of
        trajectories in each sub-ensemble is denoted by $N$.}
    \label{fig:si-excitation-separated-trpes}
\end{figure}

\section{Time and eBE windows of the TRPES regions}
\label{sec:si-trpes-signal-windows}

A common set of simulated windows was used for the spectra obtained from
squared Dyson norms, plane-wave matrix elements, and Coulomb-wave matrix
elements.

\begin{table}[H]
    \centering
    \caption{Time and electron-binding-energy windows used to delineate regions
        A--L in the experimental and simulated TRPES.}
    \label{tab:si-trpes-signal-windows}
    \begin{tabular}{
            c
            S[table-format=-2.0,table-number-alignment=right]
            @{\thinspace\textendash\thinspace}
            S[table-format=3.0,table-alignment-mode=none,table-number-alignment=left]
            S[table-format=2.1,table-number-alignment=right]
            @{\thinspace\textendash\thinspace}
            S[table-format=2.1,table-alignment-mode=none,table-number-alignment=left]
            S[table-format=-2.0,table-number-alignment=right]
            @{\thinspace\textendash\thinspace}
            S[table-format=3.0,table-alignment-mode=none,table-number-alignment=left]
            S[table-format=2.1,table-number-alignment=right]
            @{\thinspace\textendash\thinspace}
            S[table-format=2.1,table-alignment-mode=none,table-number-alignment=left]
        }
        \hline
               & \multicolumn{4}{c}{experiment}      & \multicolumn{4}{c}{simulation}                                             \\
        signal & \multicolumn{2}{c}{time / \si{\fs}} & \multicolumn{2}{c}{eBE / \si{\eV}}
               & \multicolumn{2}{c}{time / \si{\fs}} & \multicolumn{2}{c}{eBE / \si{\eV}}                                         \\
        \hline
        A      & -4                                  & 33                                 & 2.8  & 4.4  & -4  & 42  & 2.5  & 4.1  \\
        B      & 5                                   & 45                                 & 4.4  & 6.2  & 5   & 55  & 4.1  & 6.0  \\
        C      & 0                                   & 23                                 & 7.1  & 7.8  & 0   & 27  & 6.8  & 7.5  \\
        D      & 23                                  & 66                                 & 7.1  & 8.5  & 27  & 66  & 6.9  & 8.2  \\
        E      & 30                                  & 66                                 & 9.3  & 10.0 & 30  & 66  & 9.1  & 9.8  \\
        F      & 30                                  & 66                                 & 10.5 & 11.9 & 30  & 66  & 10.4 & 11.9 \\
        G      & -13                                 & -3                                 & 8.9  & 10.4 & -13 & -3  & 8.8  & 10.3 \\
        H      & 0                                   & 15                                 & 9.3  & 9.9  & 0   & 15  & 9.2  & 9.8  \\
        I      & -6                                  & 15                                 & 10.6 & 11.9 & -6  & 15  & 10.6 & 11.9 \\
        J      & 66                                  & 214                                & 7.8  & 8.7  & 66  & 214 & 7.4  & 8.4  \\
        K      & 66                                  & 214                                & 9.1  & 10.2 & 66  & 214 & 9.0  & 10.0 \\
        L      & 66                                  & 214                                & 10.5 & 11.9 & 66  & 214 & 10.4 & 11.9 \\
        \hline
    \end{tabular}
\end{table}

\section{Electronic-state and ionization-energy scans}
\label{sec:si-state-ionization-scans}

The electron-binding energies were obtained as the energy differences between
the cationic and neutral states at each interpolated geometry.

\begin{figure}[H]
    \centering
    \includegraphics[width=0.95\linewidth]{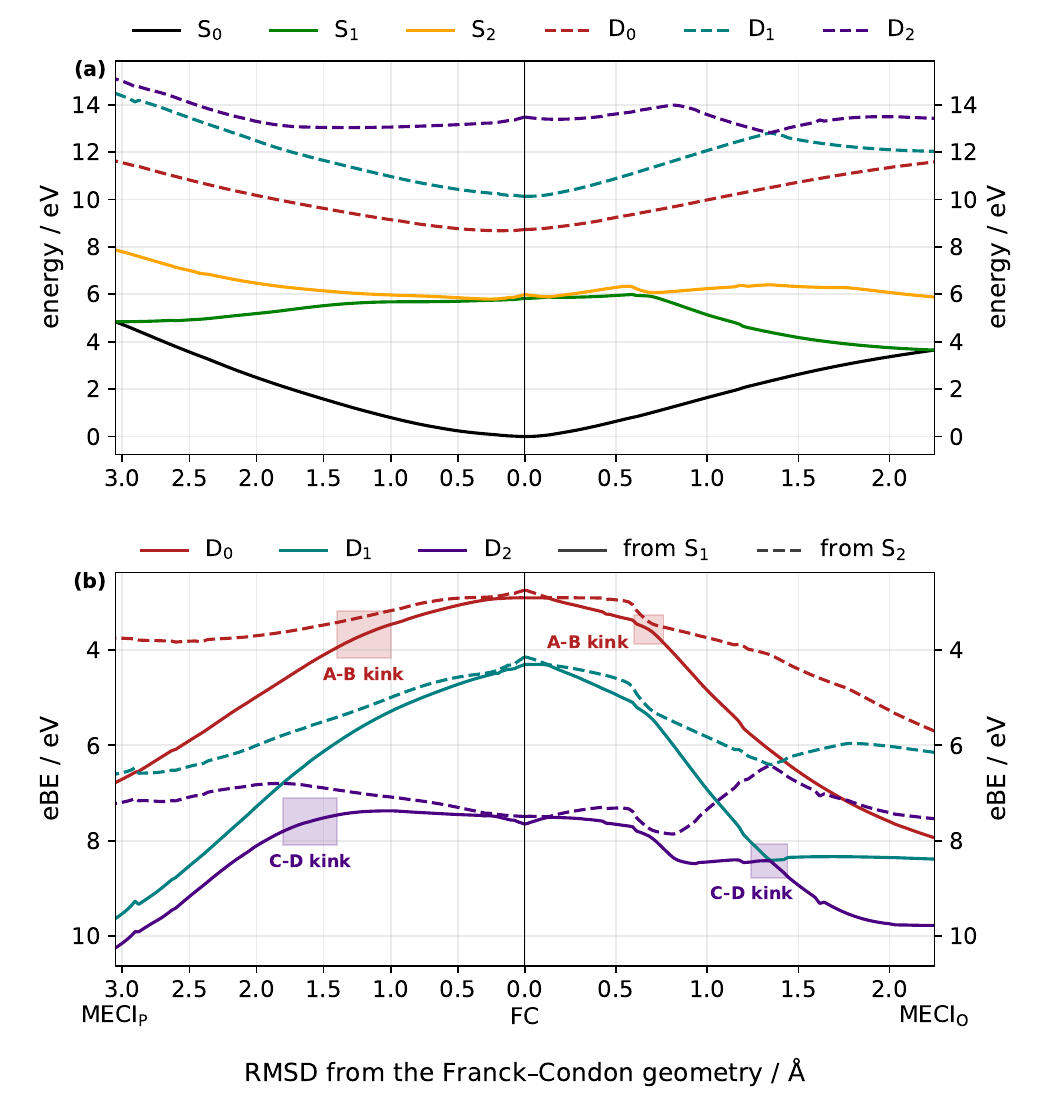}
    \caption{Electronic-state and ionization-energy scans along the interpolated
        paths from the Franck--Condon geometry toward the ring-puckering
        (MECI\textsubscript{P}) and ring-opening (MECI\textsubscript{O}) minimum-energy
        conical intersections. (a) Energies of the neutral states
        S\textsubscript{0}--S\textsubscript{2} (solid lines) and cationic states
        D\textsubscript{0}--D\textsubscript{2} (dashed lines), relative to the
        neutral ground-state energy at the Franck--Condon geometry. (b)
        Electron-binding energies for ionization from S\textsubscript{1} (solid
        lines) and S\textsubscript{2} (dashed lines). Color denotes the final
        cationic state. The red boxes mark the A--B kink, while the purple boxes
        mark the C--D kink.}
    \label{fig:si-state-ionization-scans}
\end{figure}

\clearpage
\section{Ionization-channel strengths along the MECI pathways}
\label{sec:si-dyson-norm-scans}

Because the values are evaluated along geometry scans, their variation is
independent of the trajectory populations. In addition to the
S\textsubscript{1}$\to$D\textsubscript{1} channel discussed in the main
manuscript, the corresponding channel from S\textsubscript{2} also gains
strength at distorted geometries.

\begin{figure}[H]
    \centering
    \includegraphics[width=0.95\linewidth]{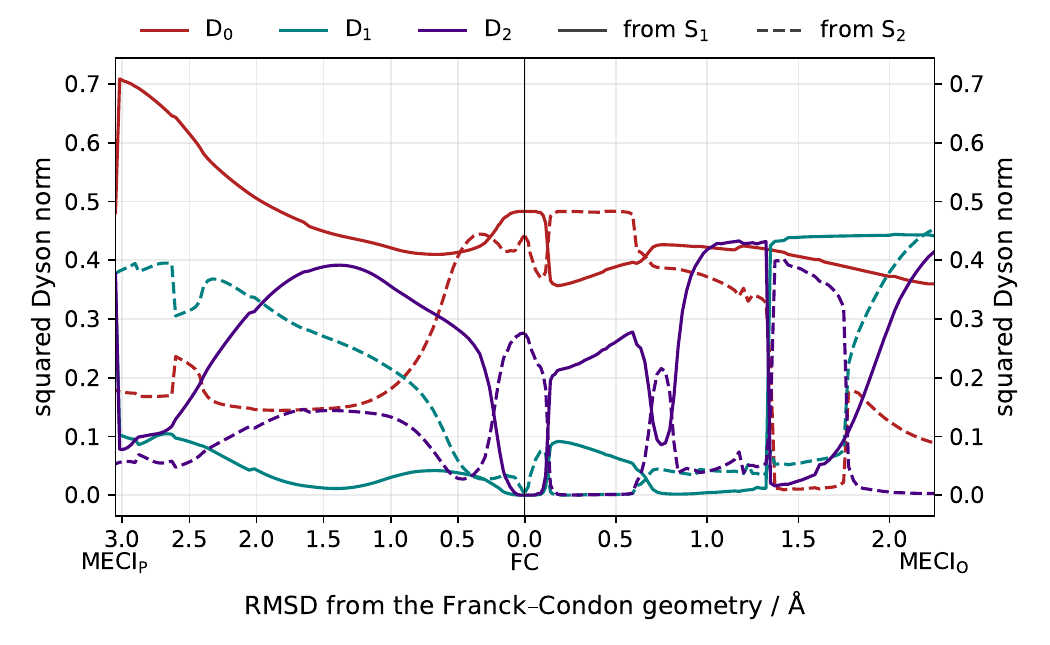}
    \caption{Squared Dyson norms along the interpolated paths from the
        Franck--Condon geometry toward MECI\textsubscript{P} and
        MECI\textsubscript{O}. Solid and dashed lines denote ionization from
        S\textsubscript{1} and S\textsubscript{2}, respectively, and color
        denotes the final cationic state.}
    \label{fig:si-dyson-norm-scans}
\end{figure}

\section{TRPES separated by final photoproduct}
\label{sec:si-product-separated-trpes}

\begin{figure}[H]
    \centering
    \includegraphics[width=0.95\linewidth]{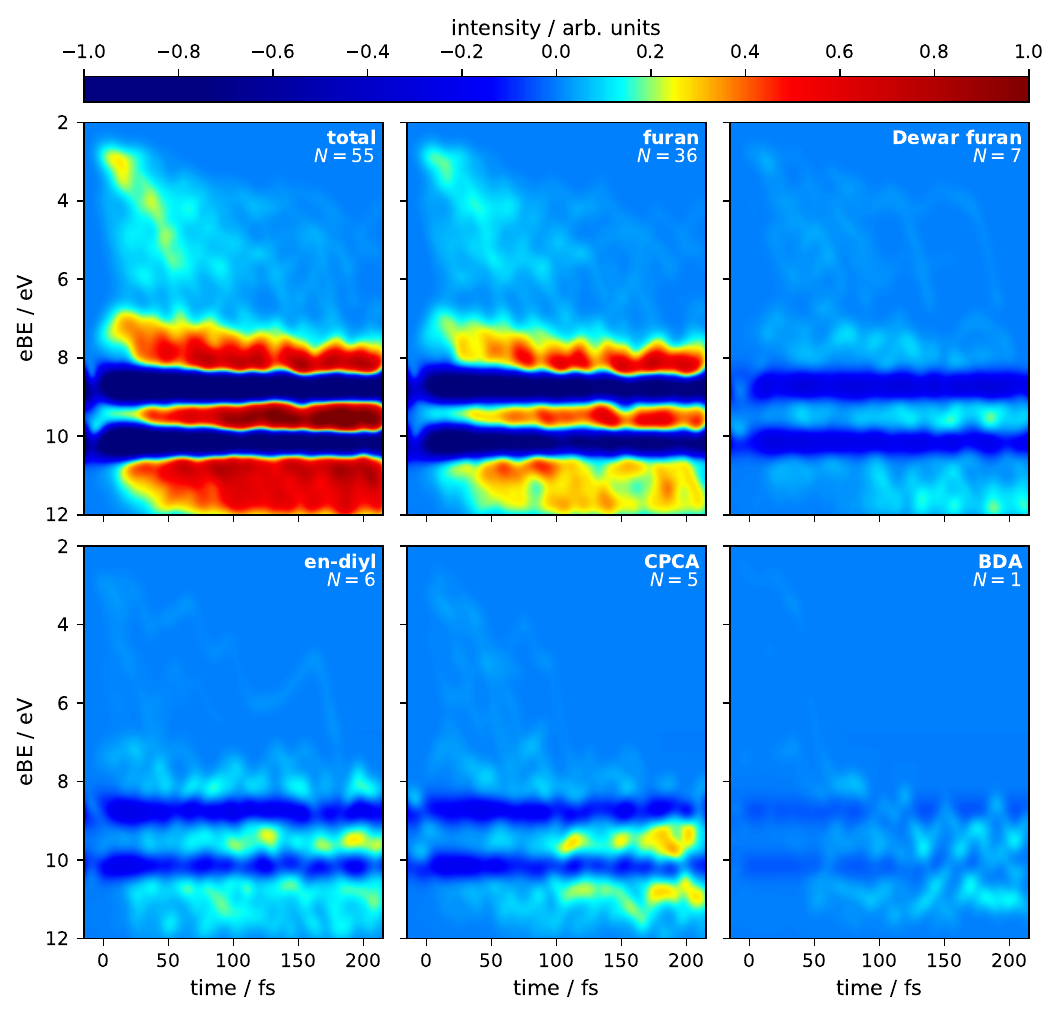}
    \caption{Simulated pre-aligned, Dyson-centered Coulomb-wave TRPES for the
        complete trajectory ensemble and separated according to the final
        photoproduct assignment. The number of trajectories in each
        sub-ensemble is denoted by $N$. The BDA panel is shown for
        completeness, but its single trajectory does not permit an
        ensemble-level interpretation.}
    \label{fig:si-product-separated-trpes}
\end{figure}

\section{Effect of rotational averaging on the simulated $\boldsymbol{\beta_2}$ maps}
\label{sec:si-trpad-averaging-comparison}

\begin{figure}[H]
    \centering
    \includegraphics[width=0.95\linewidth]{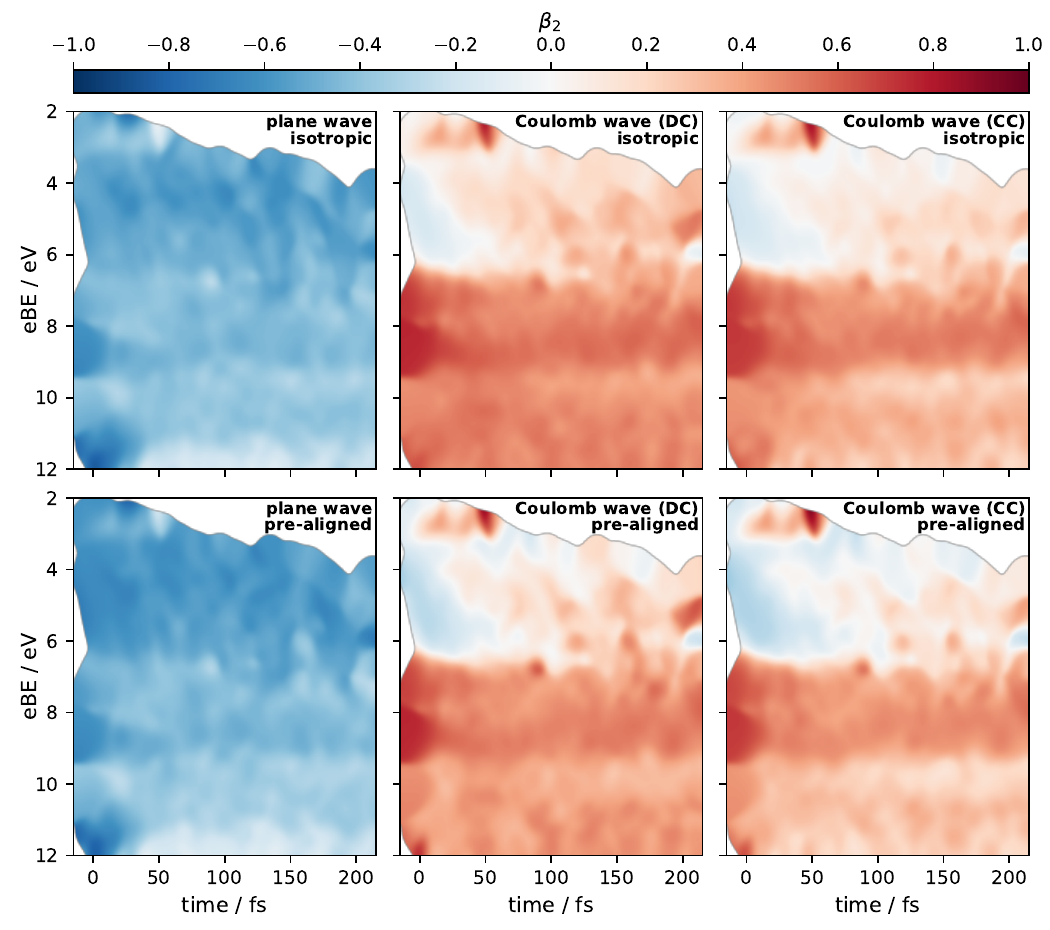}
    \caption{Comparison of simulated time-resolved photoelectron anisotropy
        parameter $\beta_2$ maps obtained using isotropic rotational averaging
        (upper row) and pre-aligned rotational averaging (lower row). The maps
        were calculated using plane-wave matrix elements (left column),
        Dyson-centered Coulomb-wave matrix elements (middle column), and
        charge-centered Coulomb-wave matrix elements (right column). Values are
        shown only where the broadened intensity exceeds $10^{-4}$ of its
        maximum.}
    \label{fig:si-trpad-averaging-comparison}
\end{figure}

\clearpage
\bibliography{bibliography,bibliography-utf}